\documentclass[10pt,journal,compsoc]{IEEEtran}
\usepackage{color}
\usepackage[hyphens]{url}  
\usepackage{graphicx} 

\usepackage{amsfonts} 
\usepackage{subfigure}
\usepackage{multirow}

\usepackage{enumitem}
\usepackage{amsmath}
\usepackage{amsthm}
\usepackage{breakurl}

\newcommand{\ie}{\emph{i.e., }}
\newcommand{\eg}{\emph{e.g., }}

\newcommand{\etc}{\emph{etc.}}
\newcommand{\wrt}{\emph{w.r.t. }}
\newcommand{\cf}{\emph{cf. }}

\ifCLASSOPTIONcompsoc
  \usepackage[nocompress]{cite}
\else
  \usepackage{cite}
\fi
\ifCLASSINFOpdf
\else
\fi
\usepackage[numbers,sort&compress]{natbib}
\usepackage{algorithmic}
\usepackage{enumitem}
\usepackage{verbatim}
\usepackage[ruled]{algorithm2e}
\usepackage[normalem]{ulem}
\useunder{\uline}{\ul}{}
\begin{document}
%
\title{Mitigating Hidden Confounding Effects for Causal Recommendation}
\author{Xinyuan Zhu, Yang Zhang, Fuli Feng, Xun Yang, Dingxian Wang and Xiangnan He	
\IEEEcompsocitemizethanks{
	\IEEEcompsocthanksitem Xinyuan Zhu, Yang Zhang, Fuli Feng, Xun Yang and Xiangnan He are with the School of Information Science and Technology, University of Science and Technology of China, Hefei, China. E-mail: zhuxinyuan@mail.ustc.edu.cn, zy2015@mail.ustc.edu.cn, fulifeng93@gmail.com, hfutyangxun@gmail.com, 	xiangnanhe@gmail.com. Dingxian Wang is with the eBay Research America, Seattle, United States. Email: diwang@ebay.com. \protect\\
	\IEEEcompsocthanksitem Fuli Feng is the corresponding author. \protect\\

}}

%
%

\markboth{Journal of \LaTeX\ Class Files,~Vol.~14, No.~8, August~2015}%
{Shell \MakeLowercase{\textit{et al.}}: Bare Demo of IEEEtran.cls for Computer Society Journals}
%



\IEEEtitleabstractindextext{%
\begin{abstract}
Recommender systems suffer from confounding biases when there  exist confounders affecting both item features and user feedback (\eg like or not).
Existing causal recommendation methods typically assume confounders are fully observed and measured, forgoing the possible existence of hidden confounders in real applications. 
For instance, product quality is a confounder since it affects both item prices and user ratings,
but is hidden for the third-party e-commerce platform due to the  difficulty of large-scale quality inspection; ignoring it could result in the bias effect of over-recommending high-price items.

This work analyzes and addresses the problem from a causal perspective.
The key lies in modeling the causal effect of item features on a user's feedback. To mitigate hidden confounding effects, it is compulsory but challenging to estimate the causal effect without measuring the confounder. 
Towards this goal, we propose a Hidden Confounder Removal (HCR) framework that leverages front-door adjustment to decompose the causal effect into two partial effects, according to the mediators between item features and user feedback. The partial effects are independent from the hidden confounder and identifiable.
During training, HCR performs multi-task learning to infer the partial effects from historical interactions. 
We instantiate HCR  for two scenarios and conduct experiments on three real-world datasets.
Empirical results show that the HCR framework provides more accurate recommendations, especially for less-active users.
We will release the code once accepted.
\end{abstract}

}

\maketitle

\IEEEdisplaynontitleabstractindextext

%
\IEEEpeerreviewmaketitle

\section{Introduction}

Data-driven models have become the default choice for building personalized recommendation services~\cite{deep-matching,SRec-survey}.
These models typically focus on the correlation between item attributes and user feedback, suffering from the confounding bias~\cite{chen_bias_2020, Saito2020AsymmetricTF}.
The source of such bias is the confounder that affects item attributes and user feedback simultaneously, leading to spurious correlations~\cite{PDA,pearl-primer}. 
For instance, the high quality of an item is the driving factor behind its high price, and it also tends to generate more positive ratings from users, resulting in a spurious correlation between high price and high rating.
Fitting the data solely based on this correlation can lead to the over-recommendation of high-price items.
Worse still, the confounding effect will hurt the fairness across item producers and make the model vulnerable to be attacked, \eg some producers may intentionally increase the price for more exposure opportunities. 
It is thus essential to mitigate the confounding effect in recommendation.

Causal recommendation has been studied to eliminate the confounding effect, which takes the causal effect of item attributes on user feedback as the recommendation criterion. 
To achieve this goal, some efforts leverage the propensity score to adjust the data distribution to be unbiased during training, aiming to obtain an unbiased model~\cite{IPS-ICML2016, Zhang2020LargescaleCA, Gruson2019OfflineET, autodebias, IpsImplicit, Guo2021EnhancedDR}. Meanwhile, others employ the do-calculus to adjust the model predictions on different values of the confounder for estimating causal effects~\cite{PDA, wenjie-kdd2021, CauSeR-CIKM2021}. These methods could effectively cut down the backdoor paths that lead to confounding effects, simulating unconfoundedness to achieve debiasing. Unfortunately, these methods cannot handle hidden confounders, since they require knowledge of the confounder's distribution to estimate the propensity score or adjust model predictions.

It is indispensable to mitigate hidden confounding effects in recommendation since many confounders are hard to measure due to technical difficulties, privacy restrictions, \etc~\cite{CausalAttention}.
For example, product quality is a such confounder in product recommendation. Most e-commerce platforms cannot monitor the productive process of items and also cannot afford the overhead to launch large-scale inspections.
News events are hidden confounders in video recommendation\footnote{Current video recommendation systems typically neglect news events~\cite{10.1145/3407190}, which are costly to be considered.}. For example, COVID-19 brings videos with face masks and attracts more user attention on epidemic-relevant videos. The spurious correlation would result in the bias of over-recommending the videos with face masks.
In food recommendation, hidden confounders can cause severe effects: some restaurants may use banned food additives (\eg poppy capsule) to please users for high ratings, but will not disclose it due to illegality; such spurious correlation may mislead the model to recommend unhealthy food. 
To mitigate such biases, it is critical to consider the hidden confounder in recommendation modeling.

\begin{figure} 
\centering
\subfigure[\textbf{ Causal graph with hidden confounders.}]{\includegraphics[width=0.18\textwidth]{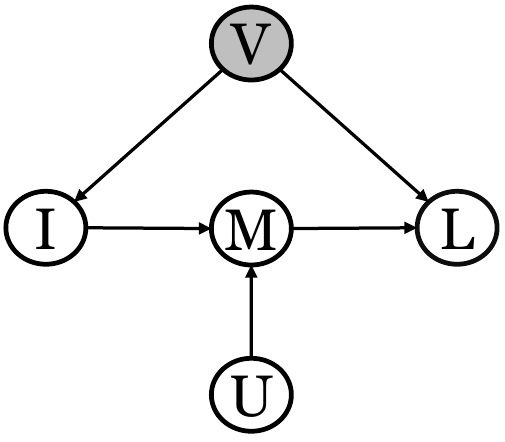}}
\subfigure[ \textbf{Intervention that cuts off the backdoor path}]{ \includegraphics[width=0.18\textwidth]{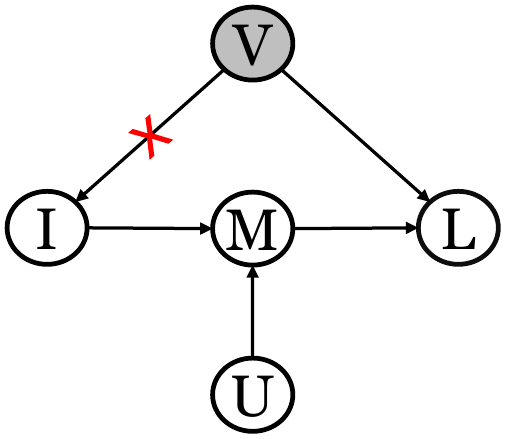}}
\vspace{-10pt}
\caption{Causal graphs for illustrating the hidden confounding effect. $V$: 
hidden confounders, $I$: item features affected by $V$, $U$: user features, $L$: user preferences, $M$: mediators. 
}
\vspace{-0.5cm}
\label{fig:causalG}
\end{figure}

Noticing the distinct properties of hidden confounder in different scenarios, we pursue a general solution for handling the hidden confounder in recommender systems. 
To understand its impact, we abstract the generation process of the \textit{like} feedback\footnote{Like conceptually denotes post-click feedback such as favorites, purchases, \etc} as a general causal graph in Figure~\ref{fig:causalG} (a). The hidden confounder ($V$) affects item features ($I$) and the happening of like ($L$) through $V\rightarrow I$ and $V\rightarrow L$, respectively. 
Item features $I$ affect $L$ through some mediators $M$ such as the interaction of user-item features and mediation feedback (\eg click).
To simplify, this work narrows its focus to scenarios where the confounder $V$ either has no influence or exerts negligible effects on the mediator $M$, \ie there is no direct edge from $V$ to $M$\footnote{In real-world scenarios, confounders that simultaneously affect both $I$ and $L$ are likely to affect $M$ to a certain degree, meaning that the edge $V \rightarrow M$ exists. The causal graph featuring the edge $V \rightarrow M$ is examined in Section \ref{Discussion_sec}.}. Under this assumption, effectively mitigating hidden confounding effects between $I$ and $L$ relies on blocking the backdoor path ($I \leftarrow V \rightarrow L$) to estimate the causal effect of $I$ on $L$, i.e., $P(L|U,do(I))$~\cite{PDA,wenjie-kdd2021}.
This is non-trivial since $V$ is unobserved, restricting us from adopting any operation requiring the value (or distribution) of the confounder $V$.

In this work, we propose a general \textit{Hidden Confounder Removal} (HCR) framework to estimate the causal effect $P(L|U,do(I))$ by performing \textit{front-door adjustment}~\cite{pearl-primer} with the assumption of no edge $V \rightarrow M$.
The core idea is to decompose the causal effect into two partial effects through the mediator $M$: 1) the effect of $M$ on $L$, \ie $P(L|U, do(M))$; and 2) the effect of $I$ on $M$, \ie $P(M|U, do(I))$). According to causal theory, both partial effects are identifiable and can be derived from plain conditional probabilities $P(M|U,I)$ and $P(L|I,M)$. In this light, we design HCR as a multi-task learning framework that simultaneously learns the two distributions from historical interactions.
After training, we infer the partial effects and chain them up to obtain $P(L|U,do(I))$, which is used for making recommendations.
We select two recommendation scenarios, e-commerce products and micro-videos, and instantiate HCR over MMGCN~\cite{MMGCN}, a representative multimodal recommendation model. We conduct extensive experiments on three real-world datasets, validating the effectiveness of HCR, especially for less-active users.

Main contributions of this work are summarized as follows:
\begin{itemize}[leftmargin=*]
    \item We study a important but rarely discussed problem of hidden confounder in recommender systems and analyze it from a causal perspective. 
    \item We propose a new causal recommendation framework, Hidden Confounder Removal, which mitigates the hidden confounding effect with front-door adjustment.
    \item We evaluate HCR in two practical scenarios and conduct extensive experiments on three real-world datasets, verifying the effectiveness of our proposal.
\end{itemize}
\section{Task Formulation}

We first give a brief introduction of notations used in this paper. We use upper characters (\eg $I$), lowercase characters (\eg $i$), and calligraphic font (\eg $\mathcal{I}$) to denote random variables, values of a random variable, and the sample space of variables, respectively. 
Taking $I$ as an example, we denote the probability distribution of a variable as $P(I)$ where the probability of observing $I=i$ from the distribution is denoted as $P(i)$ or $P(I=i)$.

From the probabilistic perspective, the target of recommendation is to estimate $P(L=1|u,i)$, which denotes the like probability between a user-item pair $(u, i)$~\cite{Wu_survey_neural}. Conventional data-driven methods parameterize the target distribution as a recommender model $f_{\Theta}(u,i)$ where $\Theta$ denotes model parameters. These methods learn model parameters from a set of historical interactions $\mathcal{D}=\left\{\left(u, i, l_{u,i}\right)|u \in \mathcal{U}, i \in \mathcal{I} \right\}$. $l_{u,i} \in \{0,1\}$ indicates the happening of like between the user $u$ and item $i$, $\mathcal{U}$ and $\mathcal{I}$ denote the user set and item set, respectively. After training, the model infers the interaction probability for each user-item pair and constructs personalized ranking accordingly.

\textit{Causal Recommendation:} To mitigate confounding biases, causal recommendation casts the recommendation problem as estimating $P(L=1|u,do(i))$, which indicates the causal effect of item features $I$ on $L$~\cite{PDA}. Note that $P(L=1|u,do(i))$ is a probability from the distribution $P(L|U, do(I))$. In the rest of this paper, we interchangeably use $P(L=1|u,do(i))$ and $P(l|u,do(i))$.
The existing work on causal recommendation estimates the causal effect $P(l|u,do(i))$ under a setting of observing all confounders between $I$ and $L$, \ie ignoring all hidden confounders. Noticing that hidden confounders are common in practice, we formulate the task as estimating the causal effect when hidden confounder exists.

\section{Approach}
In this section, we first introduce the causal graph describing the recommendation process with hidden confounders and analyze their impact.
We then present the HCR framework that aims to mitigate the hidden confounding effects, followed by an instantiation of the HCR framework.

\subsection{Causal Graph of Recommendation Process}
By definition, causal graph~\cite{pearl-primer} is a directed acyclic graph, in which a node denotes a random variable and an edge denotes a causal relation between two nodes. A causal graph describes the abstract process of data generation and can guide the modeling of causal effects~\cite{PDA,CauSeR-CIKM2021}. Figure~\ref{fig:causalG} (a) shows the generation process of the like feedback with hidden confounders. We explain the semantics of nodes and edges in the graph as follows: 
\begin{itemize}[leftmargin=*]
    \item Nodes $U$ and $I$ denote the user and item,  specifically, the corresponding user and item features.
    \item Node $L$ denotes the label of the like feedback. The like feedback conceptually denotes post-click user behaviors such as favorite, purchase, \etc
    \item Node $V$ denotes hidden confounders which affect both item features and the happening of like.
    \item Node $M$ denotes a set of variables that act as mediators between $\{U,I\}$ and $L$. 
    For example, click feedback is such a mediator, which is affected by the user and item features and is a prior behavior of the post-click feedback $L$, \ie the happening of like depends on the happening of click.
    \item Edges $I \leftarrow V \rightarrow L$ denote that $V$ affects both item features and the happening of like\footnote{As an initial attempt of studying the hidden confounder issue in recommender systems, we omit the confounders between $U$ and $L$ and the edge $V \rightarrow U$, which are left for future exploration. Besides, the spurious correlation between $U$ and $L$ unlikely changes the ranking of items for a user.}.

    \item Edges $\{U, I\} \rightarrow {M} \rightarrow {L}$ denote that $U$ and $I$ usually affect like through a set of mediators, \eg the matching of user and item features. In other words, $U$ and $I$ do not result in like solely.
    Moreover, users often demonstrate multiple cascading feedback~\cite{gao2019learning}, \eg \textit{click} $\rightarrow$ \textit{add-to-cart} $\rightarrow$ \textit{purchase} (like) in e-commerce scenarios and \textit{click} $\rightarrow$ \textit{finish} $\rightarrow$ \textit{thumbs-up} (like) on micro-video platforms. Therefore, prior feedback is also a mediator between item features $\{U, I\}$ and like feedback $L$. 
\end{itemize}

\textit{Confounding effect.} Note that the hidden confounder $V$ opens the backdoor path $I\leftarrow V \rightarrow L$, bringing spurious correlations between the item feature $I$ and like $L$. 
\begin{itemize}[leftmargin=*]
    \item As to conventional recommender models  that are trained on the historical interactions, \ie observational data, they would inherit these spurious correlations, resulting in biased estimation of user preferences.
    \item While there exist causal models  to estimate the causal effect $P(l|u,do(i))$, they can only consider observed confounders, \ie neglecting the hidden confounder $V$. Consequently, the backdoor path through $V$  still brings the confounding effect to their estimation of $P(l|u,do(i))$. These methods thus also face bias issues.
\end{itemize}

\subsection{Hidden Confounder Removal Framework}
We now consider how to mitigate the hidden confounding effect through the backdoor path $I\leftarrow V \rightarrow L$ without measuring the confounders $V$.

\subsubsection{Causal Effect Recognition}
The progress of causal inference provides us a tool to handle our case with mediator between $I$ and $L$. 
The key is \textit{front-door adjustment}, which constructs the causal effect $P(l|u,do(i))$ from the underlying effects \wrt the mediator. 
The is because any change on item features $I$ can only affect the like feedback $L$ when it has changed the value of the mediator, \eg the matching between user and item features. 
Note that $P(l | u, do(i))$ means controlling input item features $I=i$ with the \textit{do-calculus}~\cite{pearl-primer}, which is shown in Figure~\ref{fig:causalG}(b). Accordingly, we can draw the joint distribution of $L$, $V$, and $M$ as,

\begin{align}\begin{aligned} \small
    P(l,v,m|u,do(i)) &=P(m|u,do(i))P(v)P(l|m,v,u).  \label{eq:HCR_joint_dist}
    \end{aligned}
\end{align}

Eq. \eqref{eq:HCR_joint_dist} holds due to the conditional independence of variables given their parent nodes. 
Summing the probabilities in Eq. (\ref{eq:HCR_joint_dist}) over $v$ and $m$ yields the target causal effect:
\begin{align}\begin{aligned} \small
    P(l|u,do(i)) &\overset{(a)}{=}  \sum  _{m}P(m|u,do(i))\sum_{v}P(v)P(l|m,v,u)\\&
    \overset{(b)}{=} \sum_m P(m|u,do(i))P(l|u, do(m))\label{eq:causal_effect}.
\end{aligned}\end{align}

Eq. \eqref{eq:causal_effect} (b) holds due to the \textit{back-door adjustment}~\cite{pearl-primer}. In Eq. \eqref{eq:causal_effect},  $P(l|u, do(m))$ denotes the causal effect of $M$ on $L$ and \\ $P(m|u,do(i))$ denotes the causal effect of $I$ on $M$. In particular, $P(l|u, do(m))$ is the probability of like happening when forcibly setting the value of mediator as $m$. $P(m|u,do(i))$ represents how likely the mediator will be set as $m$ when we choose the item feature $i$.

According to the causal graph in Figure~\ref{fig:causalG}, we can find that both $P(l|u, do(m))$ and $P(m|u,do(i))$ are identifiable. 
\begin{itemize}[leftmargin=*]
    \item As to $P(l|u, do(m))$, we can block the backdoor path $M \leftarrow I \leftarrow V \rightarrow L$ without measuring $V$. This is because controlling $V$ is equal to controlling $I$~\cite{pearl-primer}. As such, we can achieve $P(l|u, do(m))$ by conducting a \textit{back-door adjustment} over the observable item feature $I$, which is similar to the existing causal recommendation methods on observable confounders~\cite{PDA, wang_click}.
    
    \item As to $P(m|u,do(i))$, the backdoor path $I \leftarrow V \rightarrow L \leftarrow M$ is $d$-separated by the collider $L$~\cite{pearl-primer}. Therefore, $P(m|u,do(i)) = P(m|u,i)$ where $m$, $u$, and $i$ are all observable values.
\end{itemize}

We then further derive the second term $P(l|u,do(m))$,\ie $\sum_{v}P(v)P(l|m,v,u)$ as:
\begin{align}\small
\begin{aligned}\small
    \sum_{v}P(v)P(l|v,m,u) &\overset{(a)}{=} \sum_v \sum_i P(v|i)P(i) P(l|v,m,u) \\
    &\overset{(b)}{=} \sum_{i}\sum_v P(v|i)P(i)P(l|m,v,u,i)\\
						& \overset{(c)}{=} \sum_{i}\left(\sum_v P(l|m,v,u,i)P(v|i,m)\right)P(i)\\
						 &\overset{(d)}{=} \sum_{i} P(l|u,i,m)P(i).
						 \end{aligned}
						 \label{eq:reduce_v}
\end{align}
The derivation is explained step by step as follows:
\begin{itemize} [leftmargin=*]
    \item (a) holds since $P(v)=\sum_{v}P(v|i)P(i)$.
    \item (b) is based on $P(l|m,v,u)=P(l|m,v,u,i)$ since $I$ is independent with $L$ given ${V,M}$ according to the causal graph.
    \item (c) is based on $P(v|i) = P(v|i,m)$, since $M$ is independent with $V$ given $I$ according the causal graph.
    \item (d) holds due to the properties of marginal distribution.
\end{itemize}

Note that $P(l|u, i, m) = P(l|i, m)$ holds since $l$ and $u$ are conditionally independent given $i$ and $m$. By replacing $P(l|u, do(m))$ with $\sum_{i} P(l|i,m)P(i)$ and $P(m|u,do(i))$ with $P(m|u,i)$, we obtain the causal effect free from the hidden confounder $V$, which is:
\begin{equation}
  \begin{split}
    P(l|u,do(i)) = \sum_m P(m|u,i)\sum_{i'}P(l|i',m)P(i').
  \end{split}
  \label{eq:fda}
\end{equation}

Up to this point, we have freed the causal effect $P(l|u,do(i))$ from the hidden confounder $V$. We then consider estimating the causal effect from historical data $\mathcal{D}$. According to Eq.~(\ref{eq:fda}), to obtain $P(l|u,do(i))$, 
we need to: 1) in the training stage, estimate the conditional mediator probability  $P(m|u,i)$  and the conditional like probability $P(l|i,m)$ through historical data $\mathcal{D}$; 2) in the inference stage, avoid iterating over all values of $I$ and $M$ since it is computationally costly. We need to get rid of the sum over $i^{\prime}$ and $m$ in Eq. \eqref{eq:fda}.

\subsubsection{Estimation in Training Stage} 
We estimate the two conditional probabilities $P(m|u,i)$ and $ P(l|i,m)$ in the following steps:

\vspace{+3pt}
\noindent \textbf{Step 1. Modeling the conditional mediator probability} $P(m|u, i)$. 
We parameterize the distribution of the conditional mediator probability as $f_{m}(u,i)$, 
where $f(\cdot)$ can be arbitrary backbone models (\eg MMGCN) that take $u$ and $i$ as inputs, and $f_{m}$ denotes the predicted probability of $M=m$.

\noindent \textbf{Step 2. Modeling the conditional like probability} $P(l|i, m)$. 
Considering $m$ is affected by $u$ and 
$P(l|u, i, m) = P(l|i, m)$, we parameterize the distribution $P(l|i, m)$ with the variables $u$, $i$, and $m$, employing the following decomposition:
\begin{equation}\label{eq:decompose1}
h(u,i,m)= h^{1}(u,m) * h^{2}(u,i),
\end{equation}
where $h^{1}(\cdot)$ and $h^{2}(\cdot)$  can be any backbone models for recommendation. 
Similar to~\cite{PDA,wang_click}, our main consideration for the decomposition is that the correlation $P(l|i,m)$ comes from two different sources: (1) $M$ is correlated with $L$ due to the casual path $M\rightarrow L$ given $I$; (2) $I$ is correlated with $L$ due to the backdoor path $I\leftarrow V\rightarrow{L}$ given $M$.

\begin{figure}
  \centering
  \includegraphics[scale=0.5]{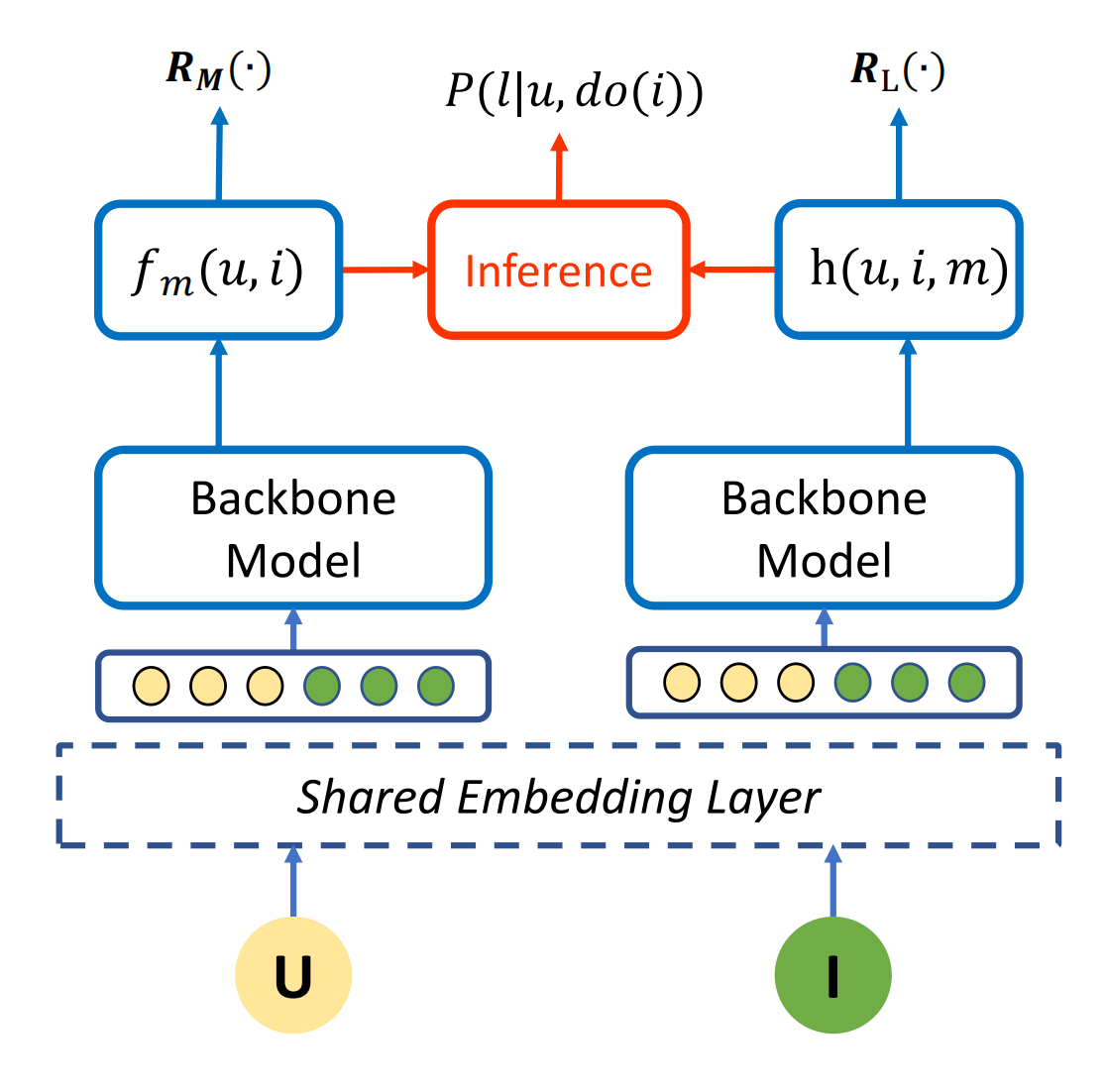}
  \vspace{-0.5cm}
  \caption{Model architecture for HCR. The training stage and inference stage use blue arrows and red arrows, respectively.}
    \vspace{-0.5cm}
  \label{fig:arc}
\end{figure}
\vspace{+3pt}
\noindent \textbf{Step 3. Estimating} $P(m|u,i)$ and  $P(l|u,i,m)$.
As the backbone models have different target values, we adopt multi-task learning to learn them simultaneously.
Formally, 
\begin{align}\label{eq:HCR-loss} \small
        \underset{f_m,h}{\min } \!\! \sum_{\left(u, i, m \right) \in \mathcal{D}} \!\!\!\! R_{M}\left(f_{m}\left(u, i\right), m \right)  
         + \beta \!\!\!\!\!\!\! \sum_{\left(u, i, m, l \right) \in \mathcal{D}} \!\!\!\!\! R_{L}\left(h\left(u, i, m\right), l \right),
\end{align}
where $R_{M}(\cdot)$ and $R_{L}(\cdot)$ denote the recommendation losses of the two tasks, respectively, such as  the cross-entropy loss, and $\beta$ is a hyper-parameter to balance the two tasks. 
Note that we let the backbone models share the embedding layer to facilitate knowledge transferring across tasks~\cite{MTL_survey}.
Figure \ref{fig:arc} shows our model architecture under the multi-task learning framework. 
Note that we merge $h^{1}(u,m)$ and $h^{2}(u,i)$ in the figure for briefness.

\subsubsection{Inference}
To construct the recommendation list for each user, we need to calculate the causal effect $P(l | u, do(i))$ for each user-item pair. It is computationally costly to direct calculate the causal effect according to Eq.~(\ref{eq:fda}) as traversing all combinations of $M$ and $I$ requires $|\mathcal{M}| * |\mathcal{I}|$ times of model inference, where $|\cdot|$ denotes the size of the sample space. 
Owing to our design of decomposing $h(u,i,m)$, we can get rid of the sum operation as:
\begin{align} \label{HCR-inference} \small
    \begin{aligned} \small
        P(l|u,do(i))&{=}\sum_m P(m|u,i)\sum_{i^{\prime}}P(l|i^{\prime},m) P(i^{\prime})\\
        & {=} \sum_m f_{m}(u,i)\sum_{i^{\prime}}h(u,i^{\prime},m) P(i^{\prime})\\
        & {=} \sum_m f_m(u,i)\sum_{i^{\prime}}h^{1}(u,m) h^{2}(u,i^{\prime}) P(i^{\prime})\\
        & {=} \sum_m f_{m}(u,i) h^{1}(u,m) * \sum_{i^{\prime}} h^{2}(u,i^{\prime}) P(i^{\prime})\\
        & {\propto} S_u \sum_m f_{m}(u,i) h^{1}(u,m)\\
    \end{aligned},
\end{align}
where $S_u = \sum_{i^{\prime}} h^{2}(u,i^{\prime})$ is a constant given $u$. 
As $S_u$ will not influence the item ranking, we can safely omit it during inference.

\vspace{-0.2cm}
\subsection{Instantiation} \label{sec:instantiation}
\begin{figure}
  \centering
    \vspace{-0.3cm}
  \includegraphics[width=0.65\linewidth]{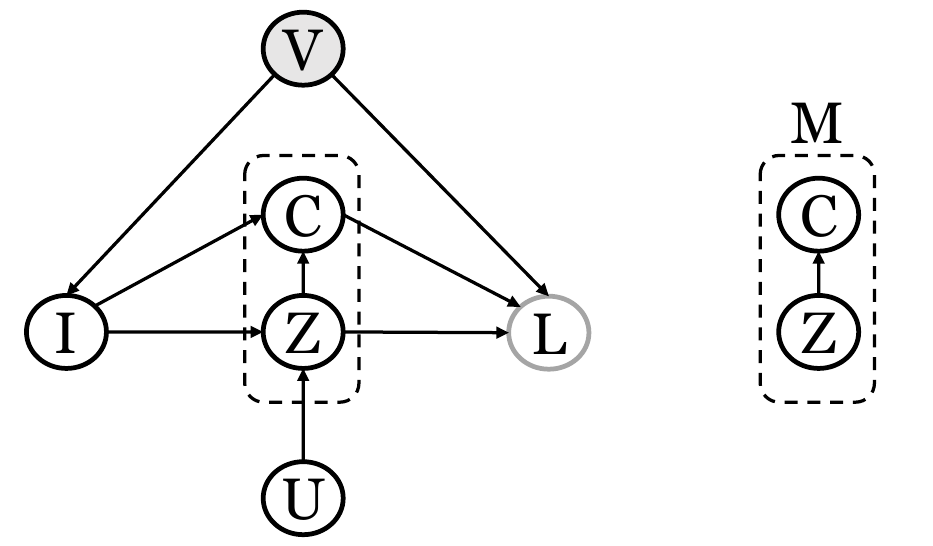}
      \vspace{-0.4cm}
  \caption{A specific causal graph under the general causal graph (Figure~\ref{fig:causalG}) where the variables $\{C,Z\}$ constitute the mediator $M$. $V$: hidden confounders, $Z$: integrated features of user-item pair, $C$: click feedback, $L$: like feedback.
  }
  \vspace{-0.5cm}
  \label{fig:CG_sp}
\end{figure}

In this part, we present an instantiation of the HCR framework in which two mediators are considered: integrated user-item features and the click feedback.
We denote the integrated features of user-item pair and the click feedback as $C$ and $Z$, respectively.
As shown in Figure \ref{fig:CG_sp}, $C$ and $Z$ also have a direct causal relation ($Z \rightarrow C$) since the integrated user-item features affect the happening of click feedback\footnote{The direct edge $I\rightarrow C$ means that some item features (\eg an eye-catching cover image) directly affects the click probability through affecting the probability of exposure~\cite{wang_click}.}.

By replacing $M$ with ${C}$ and ${Z}$ in Eq.~\eqref{eq:fda}, we identify the causal effect as:
\begin{equation}
    P(l | u, d o(i))=\sum_{c, z} P(c, z | u,i) \sum_{i^{\prime}} P\left(l |i^{\prime}, c, z\right) P\left(i^{\prime}\right). \label{eq:inst_fda}
\end{equation}
Since the like feedback are conditioned on clicks, we have $P(l=1| i^{\prime}, c = 0, z) = 0 $. Besides, the integrated feature of user-item pair, \ie $z$, is determined given the input features of  $u$ and $i$. Hence, we have $P(c, z\neq z(u,i)|u, i) = 0 $ (or $P(z\neq z(u,i)|u,i)$ = 0)~\cite{wenjie-kdd2021}, where $z(\cdot)$ represents the feature integration function in recommender systems. 
Similar to ~\cite{wenjie-kdd2021}, we implement $f_m(u,i)$ as:
\begin{equation}\label{eq:f_m_inst}
	f_m(u,i) = \begin{cases}
	\hat{f}(u,i,z(u,i)), &\text{if } z = z(u,i) \\ 
	0, &\text{else}  
		   \end{cases},
\end{equation}
where  $\hat{f}(\cdot)$ estimates $P(c=1|u,i,z)$. It corresponds to the click prediction task. $h(u,i,m)$ is instantiated by:
\begin{equation} \label{eq:h_inst}
	h(u,i,m) = \begin{cases}
	\hat{h}^{1}(u,z(u,i)) \hat{h}^{2}(u,i), &\text{if } c = 1 \\ 
	0, &\text{else}  
		   \end{cases},
\end{equation}
$\hat{h}^{1}(u,z(u,i))$, $\hat{h}^{2}(u,i)$, and $\hat{f}(u,i,z(u,i))$ are backbone recommender models, which should match the property of input features. For instance, we select MMGCN when features are in multiple modalities.
As $C$ and $L$ represents two different uesr feedback, the objective in Eq. \eqref{eq:HCR-loss} for training these backbone models is similar to multi-behavior recommendation~\cite{gao2019learning} where the hyper-parameter $\beta$ adjusts weights of different feedback.
For inference, substituting the designed models into Eq.~\eqref{HCR-inference} gives that:
\begin{equation}\label{eq:inst_P_doui}
   \begin{split}
       P(l|u,do(i)) & = S_{u} *  \hat{f}(u,i,z(u,i)) \hat{h}^{1}(u,z(u,i)) \\
    & \propto \hat{f}(u,i,z(u,i)) \hat{h}^{1}(u,z(u,i)).
   \end{split}
\end{equation}
To summarize, Algorithm~\ref{alg:HCR} shows the whole training and inference procedure of HCR under this case.

 \begin{algorithm}[t]
	\caption{Training Procedure of HCR}
	\LinesNumbered
	\label{alg:HCR}
	\KwIn{historical interactions $\mathcal{D} = \{(u_j, i_j, c_j, l_j)|u_j \in \mathcal{U}, i_j \in \mathcal{I}, c_j \in \{0,1\}, l_j \in \{0,1\}\}$ where $c_j$ and $l_j$ denote click and like label, respectively.}
	\KwOut{Identified causal effect of $i$ on $l$}
    Randomly initialize $f_m(u,i)$ and $h(u,i,m)$ as Eq. \eqref{eq:f_m_inst} and \eqref{eq:h_inst} \;
    \While{Stop condition is not reached}{
    Sample a batch $\mathcal{B}_1$ of click interactions and a batch $\mathcal{B}_2$ of like interactions from $\mathcal{D}$ \;
      Update $f_m(u,i)$ and $h(u,i,m)$ by minimizing Eq. \eqref{eq:HCR-loss};
    }
	Return causal effect $ P(l|u,do(i))$ according to Eq. \eqref{eq:inst_P_doui}.

\end{algorithm}
\vspace{-0.2cm}

\subsection{Discussion} \label{Discussion_sec}
In this part, we will discuss HCR's assumptions on hidden mediators and its general applicability considering measured confounders. Furthermore, we will investigate potential violations of the assumptions underlying HCR, with the aim of elucidating the limitations associated with its application.

\begin{figure}[t]
	\centering
        \vspace{-0.3cm}
	\subfigure[Causal graph with hidden mediator $M^\prime$]{
		\includegraphics[width=0.4\columnwidth]{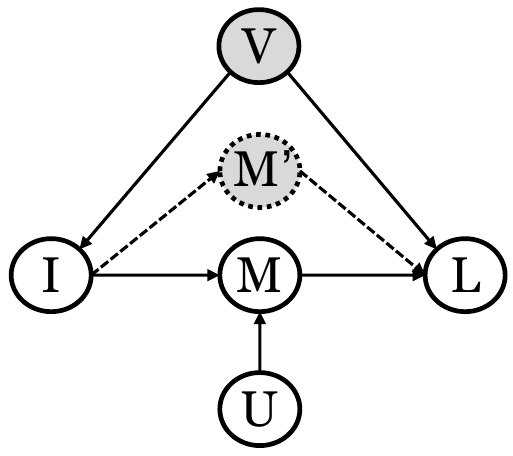}
	}
	\subfigure[Causal graph with hidden mediator $M^\prime$ and edge $V \rightarrow M^\prime$]{
		\includegraphics[width=0.4\columnwidth]{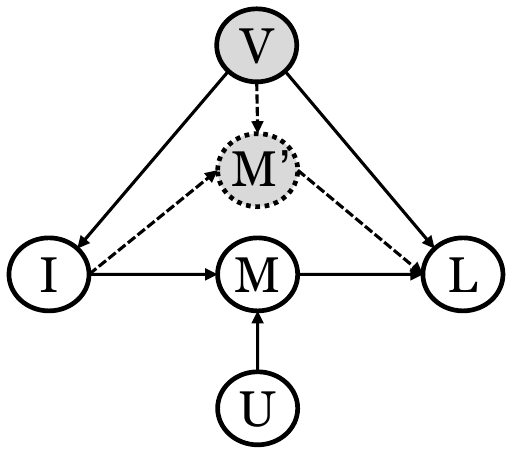}
	}
        
	\subfigure[Causal graph with measured confounder $V^\prime$]{
		\includegraphics[width=0.4\columnwidth]{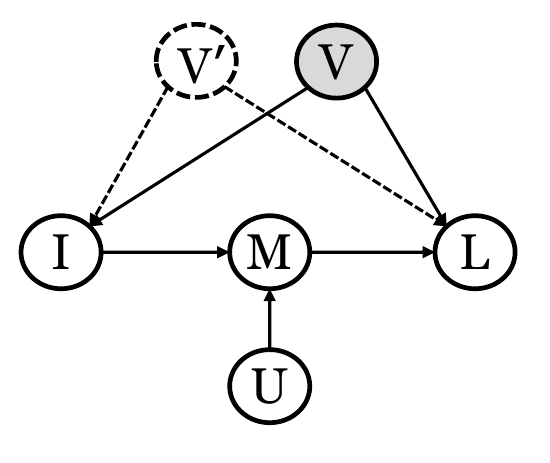} 
	}
	\subfigure[Causal graph with edge $V \rightarrow M$]{
		\includegraphics[width=0.4\columnwidth]{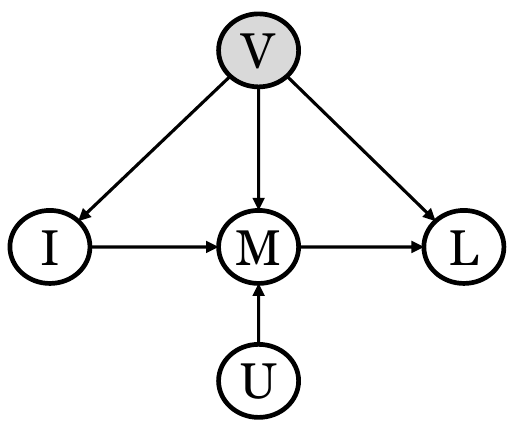} 
	}
	\vspace{-0.3cm}
	\caption{(a) Causal graph with hidden mediator $M^\prime$. (b) Causal graph with hidden mediator $M^\prime$ and edge $V \rightarrow M^\prime$. $M$ denotes the mediator that is related to the matching process of recommendation and $M^\prime$ denotes hidden mediator that is independent to the user features. (c) A special causal graph with the measured confounder $V^\prime$. $V$ denotes the hidden confounder and $V^\prime$ denotes the measured confounder.(d) A special causal graph with edge $V \rightarrow M$}
	\label{fig:discussion}
	\vspace{-0.6cm}
\end{figure}

\textit{Hidden mediators.}
Hidden mediators between $I$ and $L$ conflict with the front-door criterion, \ie HCR is under a no hidden mediator assumption. 
Generally, there could be two kinds of mediators between $I$ and $L$: 1) $M$ that is affected by user features; and 2) $M^\prime$ that is independent to user features, as shown in Fig.~\ref{fig:discussion} (a) and (b).
As to $M$, HCR accounts for both the integrated features and prior behavior of like, which should be sufficient to avoid hidden confounders. 
As to $M^\prime$, there could be two cases: 1) $M^\prime$ that is not directly affected by $V$, as shown in Fig.~\ref{fig:discussion} (a); and 2) $M^\prime$ that is directly affected by $V$ (\ie edge $V \rightarrow M^\prime$ in the causal graph), as shown in Fig.~\ref{fig:discussion} (b).
We acknowledge that HCR may not be able to handle this case due to its violation of frontdoor adjustment. However, we believe such case may be considered to be ignorable as practical recommender systems primarily focus on providing personalized services, while $M^\prime$  affects $L$ regardless of user features.

\textit{Measured confounders.} Confounders between item attributes and like feedback could be divided into two categories: hidden confounders and measured confounders, as shown in Fig.~\ref{fig:discussion} (c). The existence of measured confounders (\ie $V^\prime$) does not conflict with the proposed HCR framework because the causal effect of $I$ on $L$ is still determined through the mediator $M$\footnote{$I$ still affects $L$ only through the path $I\rightarrow M \rightarrow L$.}. Therefore, HCR is able to handle both categories of confounders.
HCR is not applicable in situations where confounders are measured but no mediator is present. In such cases, conventional methods designed for observed confounders~\cite{PDA, wang_click} can be utilized. 

\textit{Effects of hidden confounders on mediators.} There are two cases that hidden confounders directly affect mediators: 1) the hidden confounder directly affects the mediator that is independent to user features, as shown in Fig.~\ref{fig:discussion} (b). As aforementioned, we believe that this case is ignorable since practical recommender systems mainly focus on providing personalized services; 2) the hidden confounder directly affects the mediator that is affected by user features, as shown in Fig.~\ref{fig:discussion} (d). In practical scenarios, confounders between $I$ and $L$ are likely to affect $M$ to some extent. Therefore, there is an unblocked backdoor path $ I \rightarrow M \leftarrow V \rightarrow L$, violating the front-door criterion. 
HCR cannot be directly applied as it operates under the assumption that there is no direct edge from $V$ to $M$.
In scenarios where the hidden confounding effect between $I$ and $L$ is predominant, while the effect of $V$ on $M$ is comparatively weaker, HCR has the potential to mitigate a portion of the confounding effects.
However, the application of HCR is limited in scenarios where hidden confounders have non-negligible effects on $M$. In our future work, we plan to explore methods that involve more sophisticated techniques, such as instrumental variable analysis, to address this limitation.

\section{Experiments}
We conduct experiments to answer three main research questions: 

\begin{itemize}[leftmargin=*]
    \item \textbf{RQ1}: Does removing hidden confounding effects with HCR benefit the recommendation performance? How is the performance of HCR compared with existing state-of-the-art methods? 
    
    \item \textbf{RQ2}: How do the causal effect identification and components of HCR influence the effectiveness of HCR?
    
    \item \textbf{RQ3}: Where do the improvements of HCR come from, and can HCR obtain unbiased user preference estimations and achieve stable improvements?
\end{itemize}

\subsection{Experimental Settings}
\subsubsection{Datasets}
We conduct experiments on three  publicly available real-world datasets: Tiktok, Kwai, and Taobao. All datasets have multi-behaviors, one of which is the mediator -- the click feedback. The statistics of datasets are in Table~\ref{stat_dataset}.

\begin{itemize}[leftmargin=*]
    \item \textbf{Tiktok}\footnote{http://ai-lab-challenge.bytedance.com/tce/vc/.}. This is a multimodal micro-video dataset released in the ICME Challenge 2019.  It records several user feedback on videos, including \textit{click}, \textit{finish} and \textit{thumbs-up}. We view both the \textit{finish} and \textit{thumbs-up} as the like feedback, \ie $L$ in Figure~\ref{fig:CG_sp}. Tiktok dataset contains textual, visual and audio  features for items. According to \cite{wang_click}, the video captions, \ie textual features in the dataset are exposure features, which have direct effect on the clicks. 
    \item \textbf{Kwai\footnote{https://www.kuaishou.com/activity/uimc.}}. This is also a micro-video dataset released in the Kuaishou User Interest Modeling Challenge. This dataset includes three types of user feedback: \textit{click}, \textit{finish}, and \textit{thumbs-up}. Similarly, we treat \textit{finish} and \textit{thumbs-up} as  the like feedback. Besides the interaction data, the visual features of video covers were extracted by the organizer and can be viewed as exposure features. 
    \item \textbf{Taobao\footnote{\url{https://tianchi.aliyun.com/competition/entrance/231532/information}}}. This is an e-commercial dataset including users' \textit{click} and \textit{purchase} records. Besides interactions, the seller and category features are also provided. For this dataset, the \textit{purchase} will be treated as the like feedback.
\end{itemize}
We sort click feedback chronologically. The earlier 70\% clicks are used for training. For each user, liked items not included in the training set are equally divided into validation and test set, keeping chronological order. The main consideration  is that the hidden confounding effects may drift over time (\eg news events). Thus, achieving better performance on validation and test sets indicates the better capacities of recommender models to get rid of the confounding effects (in training set) and provide more accurate recommendation results. We take the like feedback to evaluate performance of all models.
\begin{table}[]
\centering
\caption{Statistics of three real-world datasets.}
\vspace{-0.3cm}
\begin{tabular}{l|l|l|l|l}
\hline
Dataset & \#Users & \#Items & \#Clicks & \#Likes\\ \hline
Tiktok  & 39,275   & 37,482   & 1,767,525  & 571,619            \\
Kwai    & 36,261   & 26,094   & 2,401,852  & 1,241,489           \\
Taobao  & 49,251   & 53,141   & 1,515,055  & 195,744 \\ \hline     
\end{tabular}
\vspace{-12pt}
\label{stat_dataset}
\end{table}

\begin{table*}[] 
\centering
\caption{Top-N recommendation performances 
on Tiktok, Kwai and Taobao. RI denotes the relative improvement of HCR compared with the best result of baselines. The best results are highlighted in bold and sub-optimal results are underlined.}
\vspace{-0.3cm}
\begin{tabular}{c|cccc|cccc|cccc}
\hline
Dataset    & \multicolumn{4}{c|}{Tiktok}                                                                
& \multicolumn{4}{c|}{Kwai}                                                                 
& \multicolumn{4}{c}{Taobao}                                                                     \\
Metric     & R@50            & \multicolumn{1}{c|}{N@50}            & R@100           & N@100           & R@50           & \multicolumn{1}{c|}{N@50}            & R@100           & N@100           & R@50             & \multicolumn{1}{l|}{N@50}             & R@100            & N@100            \\ \hline

CT         & 0.0181          & \multicolumn{1}{c|}{0.0068}          & 0.0269         & 0.0088          & 0.0158         & \multicolumn{1}{l|}{0.0072}          & 0.0270           & 0.0103          & {0.0319 }           & \multicolumn{1}{l|}{ {\ul0.0108} }           & {\ul 0.0444}           & {\ul 0.0131}           \\

CR         & 0.0209          & \multicolumn{1}{c|}{{0.0088}}          & {\ul 0.0318}          & {0.0112}          & {\ul 0.0164}         & \multicolumn{1}{l|}{ {\ul 0.0076}}          & {\ul 0.0282}          & {\ul 0.0108}          & - & \multicolumn{1}{c|}{-} & - & - \\

ESMM       & {0.0223} & \multicolumn{1}{l|}{0.0074}          & 0.0282          & 0.0087          & 0.0068         & \multicolumn{1}{l|}{0.0034}          & 0.0100          & 0.0043          & 0.0305           & \multicolumn{1}{l|}{{0.0105}}           & 0.0416           & 0.0125           \\

Multi-IPW  & {0.0229}          & \multicolumn{1}{l|}{0.0095}          & 0.0252          & 0.0100          & 0.0093         & \multicolumn{1}{l|}{0.0040}          & 0.0167          & 0.0063          & 0.0201           & \multicolumn{1}{l|}{0.0076}           & 0.0311           & 0.0097           \\

Multi-DR   & {0.0228} & \multicolumn{1}{l|}{0.0078}          & 0.0278          & 0.0090          & 0.0154         & \multicolumn{1}{l|}{0.0071}          & 0.0203          & 0.0081          & 0.0141           & \multicolumn{1}{l|}{0.0064}           & 0.0169           & 0.0069           \\ 

DCF   & \textbf{0.0245} & \multicolumn{1}{l|}{\ul 0.0102}          & 0.0297         & {\ul 0.0114}          & 0.0153         & \multicolumn{1}{l|}{0.0072}          & 0.0264          & 0.0102         & {\ul {0.0321}}           & \multicolumn{1}{l|}{0.0106}           & {\ul 0.0444}           & 0.0129           \\ \hline

HCR       & {\ul 0.0244}          & \multicolumn{1}{l|}{\textbf{0.0108}} & \textbf{0.0355} & \textbf{0.0132} & \textbf{0.0172} & \multicolumn{1}{l|}{\textbf{0.0077}} & \textbf{0.0291} & \textbf{0.0110} & \textbf{0.0350}  & \multicolumn{1}{l|}{\textbf{0.0129}}  & \textbf{0.0516}  & \textbf{0.0160}  \\

RI &       -0.41\%          & \multicolumn{1}{l|}{5.88\%}                &  11.64\%               &  15.79\%               &        

4.88\%        & \multicolumn{1}{l|}{1.32\%}                &    3.19\%             &     1.85\%            &    

9.03\%              & \multicolumn{1}{l|}{19.44\%}                 &       16.22\%           &      22.14\%   \\ \hline
\end{tabular}
\vspace{-5pt}
\label{tab:overall}
\end{table*}

\subsubsection{Baselines}
To evaluate the validity of our proposal, we compare HCR with various recommender methods, which could be categorized into two groups: normal methods (CT and ESMM) and de-biasing methods (CR, DCF, Multi-IPW and Multi-DR). These baselines are described as follows.
\begin{itemize}[leftmargin=*]
    \item \textbf{CT}~\cite{MMGCN}. This method is conducted in clean training (CT) setting, where only like feedback are used to train backbone models through a recommendation loss. it is a single-task method without considering bias issues. 

    \item \textbf{ESMM}~\cite{ma_entire_2018}. ESMM uses supervised Click-Through Rate (CTR) and Click-Through \& Conversion Rate (CTCVR) prediction tasks to implicitly train the Conversion Rate (CVR) model in a multi-task learning manner. Thus, ESMM is trained in the entire exposure space and tries to remedy selection bias and data sparsity issues in CVR estimation. ESMM is included for comparison since it also employs multi-task learning. 
    The original implementation of ESMM adopts multi-layer perception as the backbone. For a fair comparison, we replace MLP in ESMM with MMGCN to make better use of multimodal features.

    \item \textbf{CR}~\cite{wang_click}. 
    CR is a counterfactual inference-based method that addresses the clickbait issue. CR aims to capture unbiased user preferences without using like feedback. To achieve this goal, CR requires identifying exposure features, which directly affect users' clicks. Since the Taobao dataset does not contain exposure features, we only evaluate CR on Tiktok and Kwai datasets. We use code released by the authors to re-implement experiments, where CR is also implemented based on MMGCN. The hyper-parameter $\alpha$ to control the influence of exposure feature during training is tuned in range of $\{1,2,3,5\}$.

    \item \textbf{Multi-IPW}~\cite{Zhang2020LargescaleCA}. 
    It is a de-biasing method which applies the Inverse Propensity Weighting method under a multi-task learning framework. It tries to solve the selection bias issue for post-click conversion rate (CVR) estimation. For this goal, it introduces an auxiliary CTR task to remedy the data sparsity issue. Meanwhile, predictions of the CTR model are treated as propensities for the CVR task. We implement its CTR and CVR backbone models with MMGCN to exploit multimodal features.

    \item \textbf{Multi-DR}~\cite{Zhang2020LargescaleCA}. 
    This is a Doubly Robust-based method under a multi-task (CVR and CTR) learning framework. Multi-DR makes use of the IPW like Multi-IPW and adopts an imputation model to predict estimation errors for better de-biasing. Similarly, we implement its CTR, CVR and imputation model with MMGCN.

    \item \textbf{DCF}~\cite{DCF-recsys20}. This is a method that takes hidden confounders into account. Assuming item exposures are highly related to hidden confounders, it learns an exposure model by fitting exposures. The exposure model provides substitutes for unobserved confounders and then DCF leverages the substitutes to remove the impact of hidden confounders in rating data. Since no exposure data is provided in all three datasets, we use the click and like feedback to substitute exposures and ratings, respectively.  Similarly, we implement DCF based on MMGCN for a fair comparison.

\end{itemize}
The baselines can also be categorized into two groups: methods with multi-task learning framework, including ESMM, Multi-IPW and Multi-DR and methods without multi-task learning framework, including CT, CR and DCF.

\subsubsection{Evaluation Protocols}
During evaluation, recommender models serve each user and generate a recommendation list by ranking items that do not appear in the training dataset, \ie the all-ranking protocol. Since the like feedback has better ability to indicate actual user preferences, we only treat the like feedback on validation and test sets as positive samples. To measure the top-K recommendation performance, we take two widely-used evaluation metrics: Recall@K (abbreviated as R@K), which considers whether the relevant items are retrieved within top-K positions, and NDCG@K (abbreviated as R@K) that measures relative orders among positive and negative items in the top-K recommendation list. Due to the large number of items and sparsity of the like feedback in real-world datasets, we report the results of K=50 and K=100.

\subsubsection{Hyper-parameters Settings} We optimize all models with the Adam~\cite{kingma_adam_2017} optimizer and use default mini-batch size of 1024. For Tiktok and Kwai, we search learning rate in the range of \{$1e\text{-}4$, $5e\text{-}4$, $1e\text{-}3$\}. For Taobao, we seach learning rate in the range of $\{1e\text{-}5, 1e\text{-}4, 5e\text{-}4, 1e\text{-}3\}$.
For all methods,  $L_2$ regularization coefficient is searched in the range of$\{1e\text{-}4, 1e\text{-}3, 1e\text{-}2\}$, and the settings of the backbone MMGCN follow previous work CR~\cite{wang_click}, including latent dimension, concatenation strategy, the number of GCN layer, \etc~For HCR, the weight $\beta$ in the multi-task loss function (\ie Eq.~\eqref{eq:HCR-loss}) is tuned in the range of $\{1, 2, 3, 5\}$.  
Moreover, for model selections, early stopping is adopted. Training will stop if NDCG@50 in the validation set does not increase for 10 successive epochs.

\begin{table*}[]
\caption{Recommendation performance of HCR and its four
variants with different designs being disabled or replaced in the training or inference stage. The best results are highlighted in bold.}
\vspace{-10pt}
\label{variants}
\centering
\begin{tabular}{c|cccc|cccc|cccc}
\hline
Dataset &
  \multicolumn{4}{c|}{Tiktok} &
  \multicolumn{4}{c|}{Kwai} &
  \multicolumn{4}{c}{Taobao} \\
Metric &
  R@50 &
  \multicolumn{1}{c|}{R@100} &
  N@50 &
  N@100 &
  R@50 &
  \multicolumn{1}{c|}{R@100} &
  N@50 &
  N@100 &
  R@50 &
  \multicolumn{1}{c|}{R@100} &
  N@50 &
  N@100 \\ \hline
HCR &
  \textbf{0.0244} &
  \multicolumn{1}{c|}{\textbf{0.0355}} &
  \textbf{0.0108} &
  \textbf{0.0132} &
  \textbf{0.0172} &
  \multicolumn{1}{c|}{\textbf{0.0291}} &
  \textbf{0.0077} &
  \textbf{0.0110} &
  \textbf{0.0350} &
  \multicolumn{1}{c|}{\textbf{0.0516}} &
  \textbf{0.0129} &
  \textbf{0.0160} \\
HCR-T &
  0.0222 &
  \multicolumn{1}{c|}{0.0352} &
  0.0094 &
  0.0121 &
  0.0128 &
  \multicolumn{1}{c|}{0.0224} &
  0.0059 &
  0.0085 &
  0.0339 &
  \multicolumn{1}{c|}{0.0501} &
  0.0126 &
  0.0156 \\
HCR-S1 &
  0.0229 &
  \multicolumn{1}{c|}{0.0350} &
  0.0099 &
  0.0125 &
  0.0159 &
  \multicolumn{1}{c|}{0.0277} &
  0.0072 &
  0.0104 &
  0.0212 &
  \multicolumn{1}{c|}{0.0324} &
  0.0079 &
  0.0100 \\
HCR-S2 &
  0.0088 &
  \multicolumn{1}{c|}{0.0179} &
  0.0029 &
  0.0049 &
  0.0033 &
  \multicolumn{1}{c|}{0.0049} &
  0.0016 &
  0.0021 &
  0.0242 &
  \multicolumn{1}{c|}{0.0344} &
  0.0086 &
  0.0105 \\
HCR-NS &
  0.0229 &
  \multicolumn{1}{c|}{0.0325} &
  0.0105 &
  0.0126 &
  0.0131 &
  \multicolumn{1}{c|}{0.0235} &
  0.0062 &
  0.0090 &
  0.0284 &
  \multicolumn{1}{c|}{0.0442} &
  0.0092 &
  0.0122 \\ \hline
\end{tabular}
\vspace{-10pt}
\end{table*}
\subsection{Performance Comparison (RQ1)}
In this section, we study the recommendation performance of HCR framework. We compare HCR with a variety of approaches including the biased conventional methods and de-biasing methods. The comparison result is summarized in Table~\ref{tab:overall}, where we have the following observations:
\vspace{-2pt}
\begin{itemize}[leftmargin=*]
    \item In most cases, the proposed HCR achieves distinct improvements over all baselines, showing its capacity of obtaining more accurate user preference estimations. The improvement can be attributed to the deconfounded training and inference, which remove hidden confounding effects. In addition, HCR consistently outperforms all methods that do not model hidden confounders, \ie all baselines except DCF. These findings reflect the rationality of our causal analysis of hidden confounders and validate the necessity to deal with hidden confounder issues.

    \item HCR consistently outperforms ESMM, while both these two methods adopt themulti-task learning framework. This result implies that improvements of HCR over baselines should be attributed to removing the hidden confounding effects rather than the multi-task learning.
    
    \item CR outperforms CT on Tiktok and Kwai. It is confusing that the model trained with direct access to the like (CT) cannot beat a model using only clicks (CR) during evaluation on the like data. However, this phenomenon is also found in the CR paper~\cite{wang_click}. Two main reasons exist: 1) highly sparse like data; 2) CT captures correlations between like and item features without eliminating spurious correlations brought by confounders, leading to biased estimations. CR employs counterfactual inference to eliminate biases in clicks, resulting in improved user preference estimations. Meanwhile, HCR outperforms CR \wrt R@50 by $16.75\%$ and $4.88\%$ on Tiktok and Kwai, respectively. These results again show that it is essential to address hidden confounders.

    \item Multi-IPW and Multi-DR are de-biasing recommendation methods, and they outperform CT on Tiktok dataset. However, they cannot maintain improvements across all datasets. This phenomenon may be due to the high variance of IPW-based methods~\cite{Saito2020AsymmetricTF}. Another reason is that in order to achieve the desired unbiased estimations, these two methods assume the non-existence of hidden confounders, which is usually impossible in practice~\cite{DCF-recsys20}. 
    
    \item DCF performs well in certain scenarios by controlling hidden confounding effects using substitute variables. However, DCF falls short compared to other baselines in several cases, and it is outperformed by the proposed HCR (except for Recall@50 on Tiktok). Substitute confounder estimations may not effectively handle arbitrary hidden confounders, particularly when confounders have weak correlations with exposures. In contrast, HCR directly identifies the causal impact of item features on user preferences, without explicitly measuring or estimating them. 

\end{itemize}

\vspace{-0.3cm}
\subsection{Ablation Study (RQ2)}
To shed light on performance improvements, we further study four variants of HCR, named HCR-T, HCR-S1, HCR-S2, and HCR-NS, respectively. The former three variants differ from the original HCR during inference, with components of models disabled or replaced. The HCR-NS model disables the shared embedding layer between the two estimation modules in the training stage.

    \noindent \textbf{-HCR-T}. Recall that we decompose $h_m(u,i,m)$ as the product of $h^{1}(u,z)$ and $h^{2}(u,i)$ as described in Eq. ~\eqref{eq:decompose1}. This decoupling design enables rid of the sum over $i^{\prime}$ through removing $h^{2}(u,i^{\prime})$ in the inference stage. For the variant HCR-T, we adapt the inference formula to,
    \begin{equation}
    \vspace{-0.2cm}
       \begin{split}
           P_{HCR-T} = \hat{f}(u,i,z(u,i)) \hat{h}^{1}(u,z(u,i)) \hat{h}^{2}(u,i).
       \end{split}
    \end{equation}
In fact, it solely relies on correlations to represent the effect of $M$ on $L$, disabling intervention during inference.
    
    \noindent \textbf{-HCR-S1} denotes the model in which we disable $\hat{h}^{1}(u,z(u,i))$ in the inference stage, formulated as,
    \begin{equation}
         P_{HCR-S1}  = \hat{f}(u,i,z(u,i)).  \\
    \end{equation}
    
    \noindent \textbf{-HCR-S2} denotes the model in which we disable $\hat{f}(u,i,z(u,i))$ during inference, formulated as,
    \begin{equation}
         P_{HCR-S2}  = \hat{h}^{1}(u,z(u,i)).  \\
    \end{equation}
    
    \noindent \textbf{-HCR-NS} represents that we disable the shared embedding layer between the two probability estimation models in the training stage. During inference, HCR-NS adopts the same scoring function as the original model. 

The recommendation performance of HCR and its four variants on the three datasets are summarized in Table \ref{variants}. We can obtain the following findings from the results:
\begin{itemize}[leftmargin=*]
    \item  HCR consistently outperforms HCR-S1 and HCR-S2, indicating that using a single estimation model in the HCR framework as ranking scores harms recommendation results.
    Especially, $\hat{h}^{1}(u,z(u,i))$ (adopted by HCR-S2) can be viewed as the partial effect of $M$ on $L$ while forgoing the effect of $I$ on $M$. The absence of complete causal effects explains performance drops of HCR-S1 and HCR-S2. These results confirm the effectiveness of the multi-task framework and causal recognition.

    \item During inference, HCR-T directly combines trained estimation functions fitted in the observed data to perform recommendations. Thus the ranking scores of HCR-T cannot reflect the causal effects, but only correlations, resulting in worse performance compared to HCR. Therefore, the superior performance of HCR over HCR-T reflects rationalities of causal effect identification and the capacities of HCR to mitigate hidden confounding effects. 

    \item HCR achieves consistent gains over HCR-NS in all cases. Meanwhile, ESMM that also adopts embedding sharing without causal interventions is outperformed by HCR~(\cf Table~\ref{tab:overall}). These  results imply the necessity of combining embedding sharing and causal effect identification. We attribute the performance drop of HCR-NS to its failure to facilitate knowledge transfer across tasks. Thus, the accurate causal effect $P(l|u, do(i))$ cannot be achieved due to the insufficient estimations of the required correlations.
\end{itemize}

\subsection{In-depth Analyses (RQ3)}

In this subsection, we conduct comparisons between HCR and CT as examples to further investigate:
1) where the main performance improvements come from; 2) whether the improvements are stable along time; and 3) whether HCR recommends high-quality items. 

\subsubsection{Improvements in Active and Less-active User Groups}
\begin{figure}[t]
	\centering
	\subfigure[Absolute performances.]{
		\includegraphics[width=0.5\columnwidth]{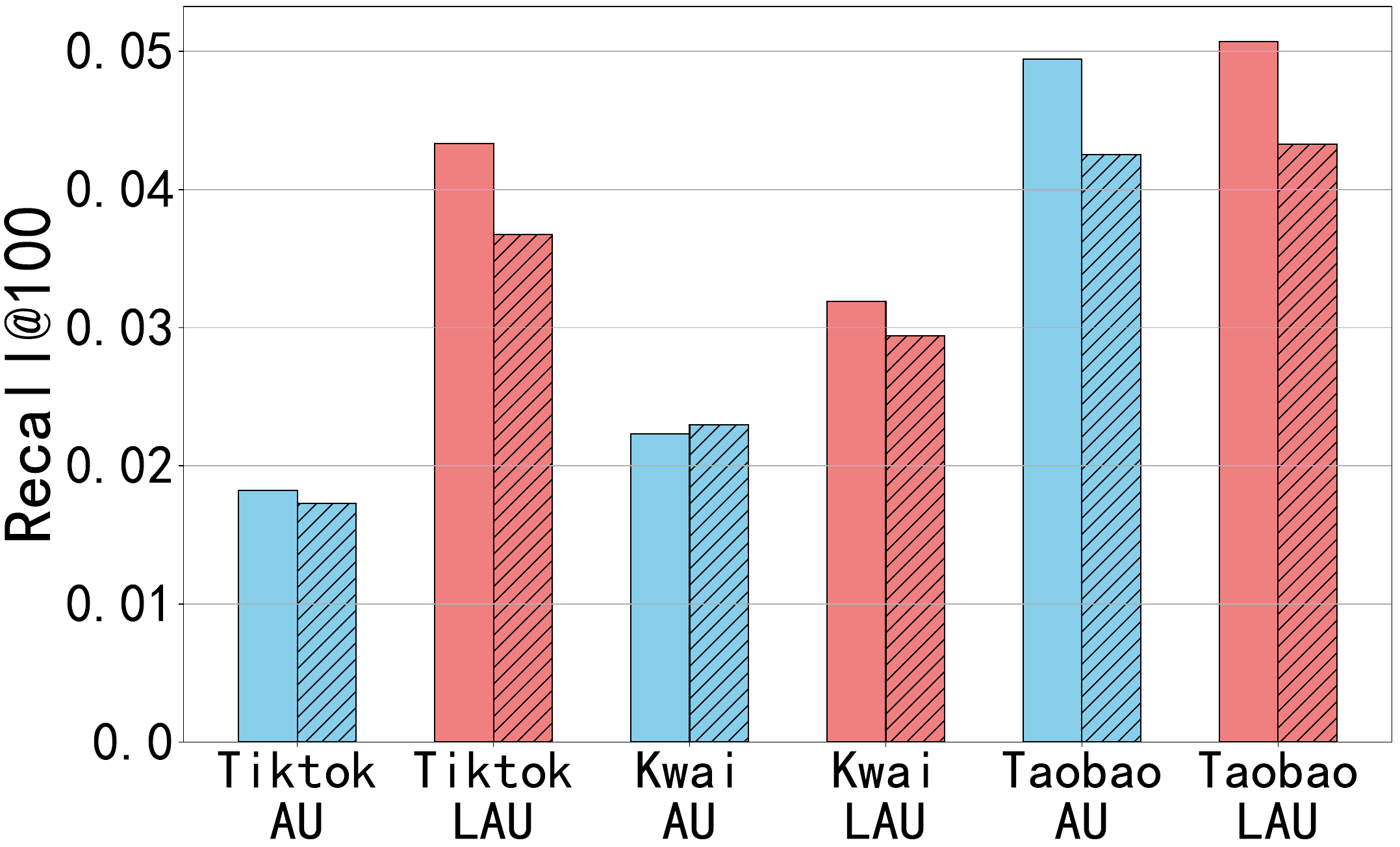}
	}
	\subfigure[Relative improvements of HCR over CT]{
		\includegraphics[width=0.36\columnwidth]{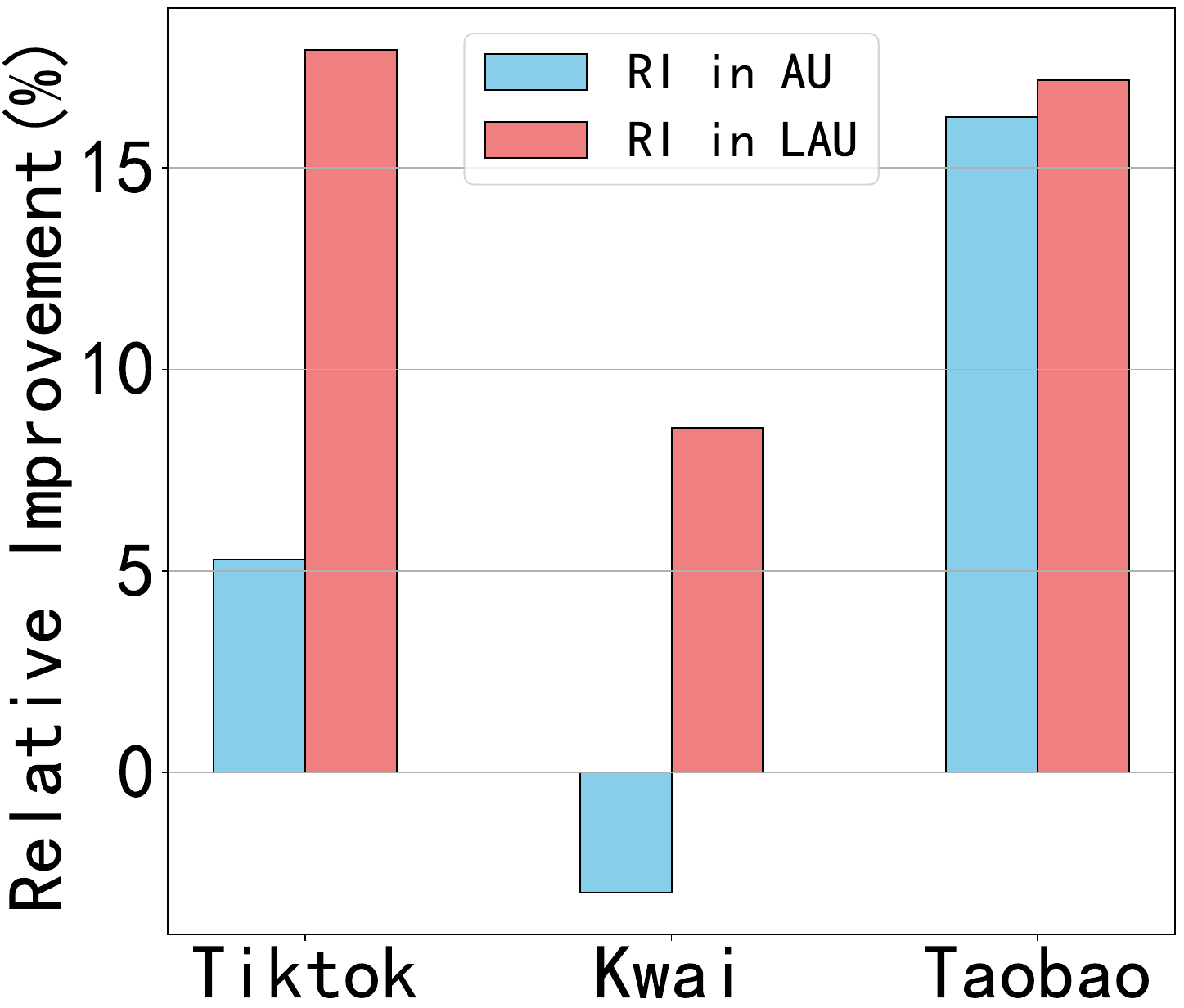}
	}
	\vspace{-0.3cm}
	\caption{Performance of HCR and CT in active and less-active user groups. (a) the absolute performance; and (b) the relative improvements of HCR over CT. "AU" and "LAU" are short for the active user group and the less-active user group, respectively. In (a), bars with slash and without slash corresponds to CT and HCR, respectively. Better viewed in color.}
	\label{fig:active}
	\vspace{-0.45cm}
\end{figure}

In this part, we examine if HCR provides greater enhancements for active or less-active user groups. For Tiktok and Kwai datasets, the top 40\% users based on click count are considered active, while the rest are considered less-active. Due to higher sparsity of Taobao dataset, we select top 20\% users as the active users.  We compare 1) the absolute performance \wrt Recall@100 of HCR and CT; 2) the relative improvements of HCR over CT. The experimental results are summarized in Figure \ref{fig:active}. 

According to Figure \ref{fig:active}, 1) both HCR and CT achieve better performance in less-active user groups. We propose that active users usually have diverse preferences, and their interests drift with time with higher probabilities. Recall that datasets are chronologically split. The inherent user interests in training and test sets are different, leading to the performance drop. In contrast, less-active users have stable interests.  2) HCR can outperform CT on both active and less active user groups in most cases (5 of 6), which verifies the superiority of HCR. 
3) HCR achieves greater relative improvements in less-active user group. As aforementioned, active users have more unstable interests. 
Even though HCR obtains relatively more precise estimations of users' interests during training, it cannot achieve greater improvements on the test set since users' interests have drifted. 
While the less active users have more stable interest, thus the unbiased estimation of HCR (regarding hidden confounders) could show greater superiority over the biased estimation of CT. Meanwhile, improving the experience of less-active users is meaningful, since a large percentage of users belong to the less-active group due to the long-tail phenomenon.

\subsubsection{Improvement along Time}
\begin{figure}[t]
	\centering
	\subfigure[Relative improvements in subsets.]{
		\includegraphics[width=0.45\columnwidth]{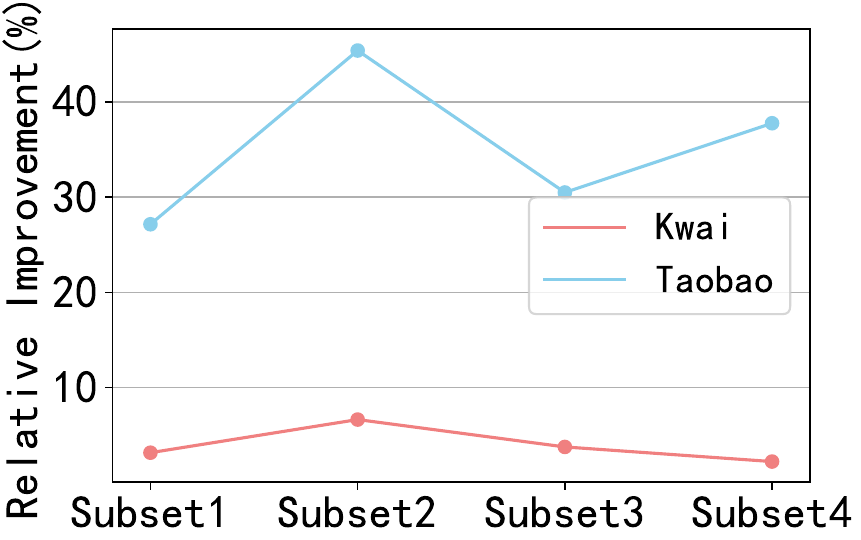}
	}
	\subfigure[Average drop of performance.]{
		\includegraphics[width=0.45\columnwidth]{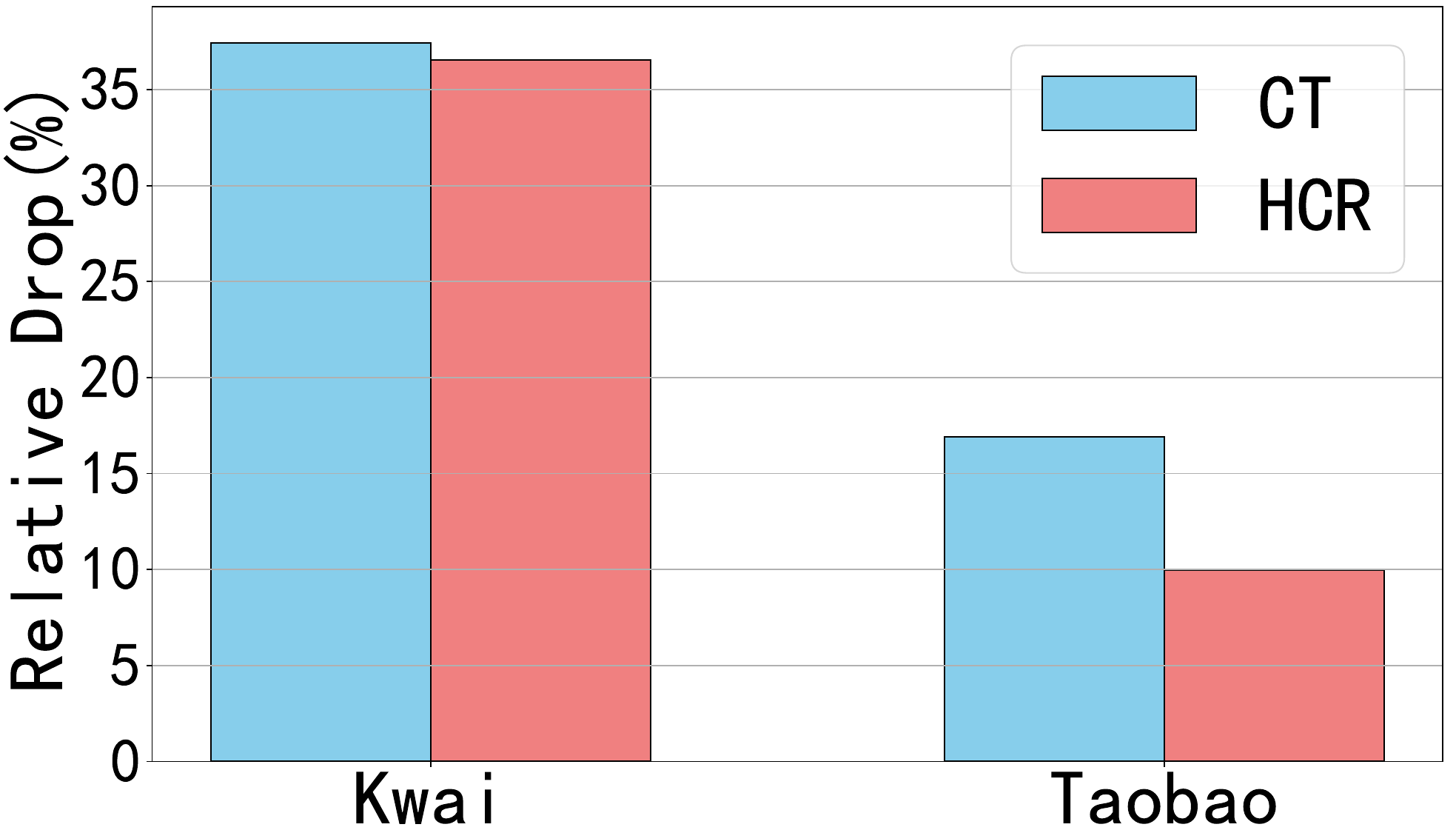}
	}
	\vspace{-0.35cm}
	\caption{(a) Relative improvements of HCR over CT in four subsets with different testing periods; (b) the average performance drop of HCR and CT on subsets 2,3,4 compared to subset 1. The four subsets are in chronological order, \ie subset 1 is the closest to the training data.}
	\label{fig:chronological}
	\vspace{-0.4cm}
\end{figure}

As aforementioned, the value of hidden confounders may change along time, \eg different types of social event occur on different dates, which means its impact drifts over time. Therefore, we evaluate the performance of HCR over time compared to CT. For each user, we evenly divide the corresponding liked items in validation and test sets into four subsets chronologically, denoted as subset 1, 2, 3, and 4 respectively. 
We conduct two experiments: 1) compare relative improvements of HCR over CT in four subsets; 2) evaluate average performance drop of HCR and CT in subsets 2, 3, and 4, compared to subset 1. The results are summarized in Figure~\ref{fig:chronological}. Figure~\ref{fig:chronological} (a) shows that in all subsets, HCR achieves consistent gains over CT. Figure~\ref{fig:chronological} (b) shows that HCR maintains a relatively small performance drop compared to CT. These findings can be attributed to the mitigation of dynamic hidden confounding effects, resulting in more accurate estimations of user preferences. However, even though HCR consistently outperforms CT, its performance still declines over time. We attribute this fact to the drift of user interest. Recommender models always tend to perform worse on subsets later in time. In the future, we may need to design models that remove the impact of hidden confounders dynamically to capture real-time unbiased user preferences.

\subsubsection{Recommendation Results \wrt Like/click Ratio}
\begin{figure}[t]
	\centering
	\subfigure[Item group with high like/click ratio.]{
		\includegraphics[width=0.45\columnwidth]{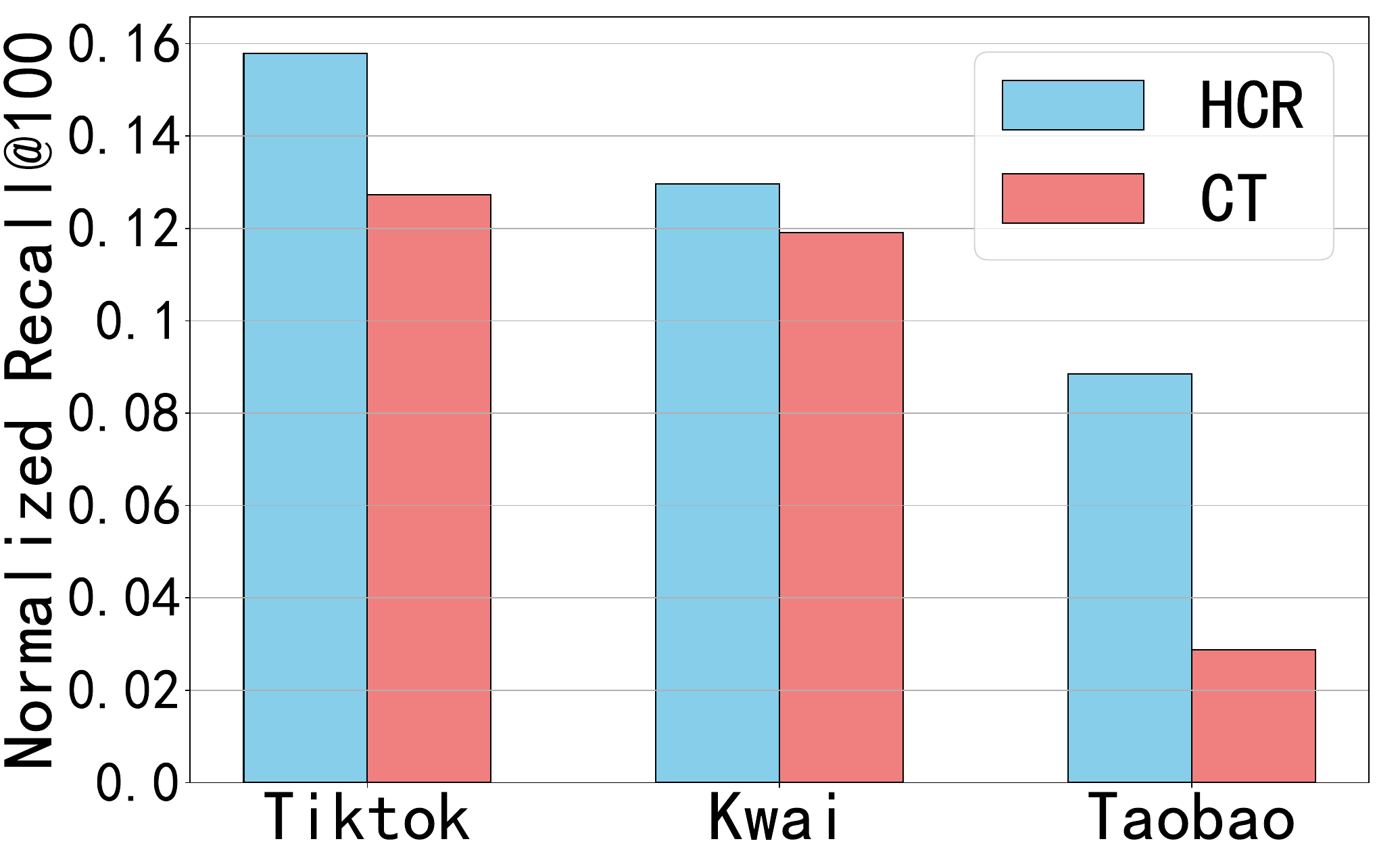}
	}
	\subfigure[Item group with low like/click ratio.]{
		\includegraphics[width=0.45\columnwidth]{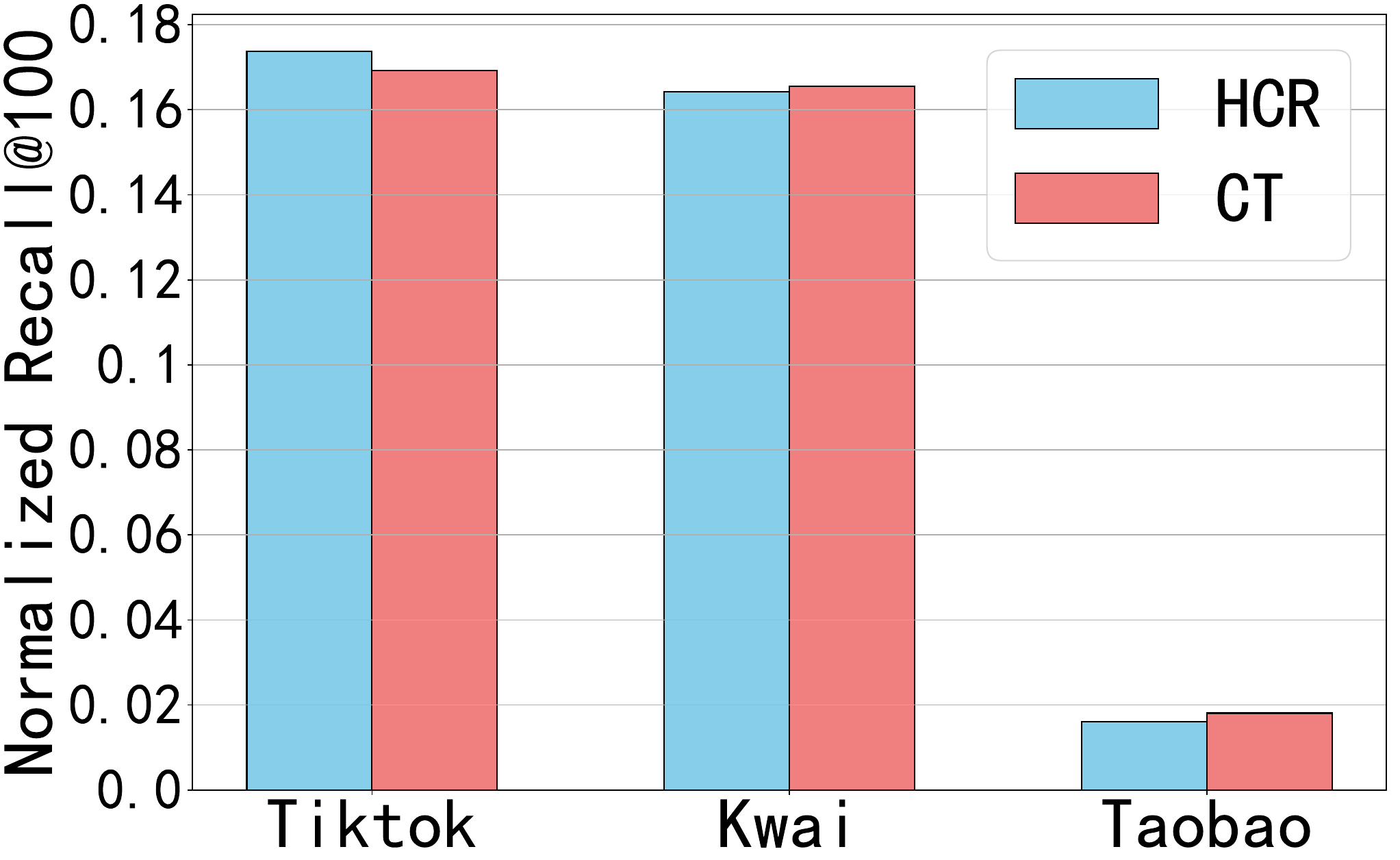}
	}
	\vspace{-0.35cm}
	\caption{Normalized Recall of HCR and CT in item groups with high and low like/click ratio.}
	\label{fig:like_click_ratio}
	\vspace{-0.4cm}
\end{figure}

Whether the proposed HCR can achieve consistent and unbiased user preference estimations is our concern. Thus in this part, we sort items according to like/click ratio and divide them into two subsets with a 1:2 ratio of size. Items with higher like/click ratios are more likely to satisfy users' interest, while over-recommending items with low like/click ratios is likely to hurt user experience. We use normalized recall to evaluate recommender models, defined as the recall metric normalized by proportions of target item group in recommendation list. The performance of HCR and CR in the two groups are shown in Figure \ref{fig:like_click_ratio}. HCR achieves consistent gains over CT in item groups with higher like/click ratios, showing that HCR can provide more recommendations with high quality items due to the removal of hidden confounders.

\subsubsection{Effects of Confounding Strength}
\begin{figure}[t]
	\centering
	\subfigure[Absolute performance of HCR and CT and relative improvements of HCR over CT with different mask ratios.]{
		\includegraphics[width=0.45\columnwidth]{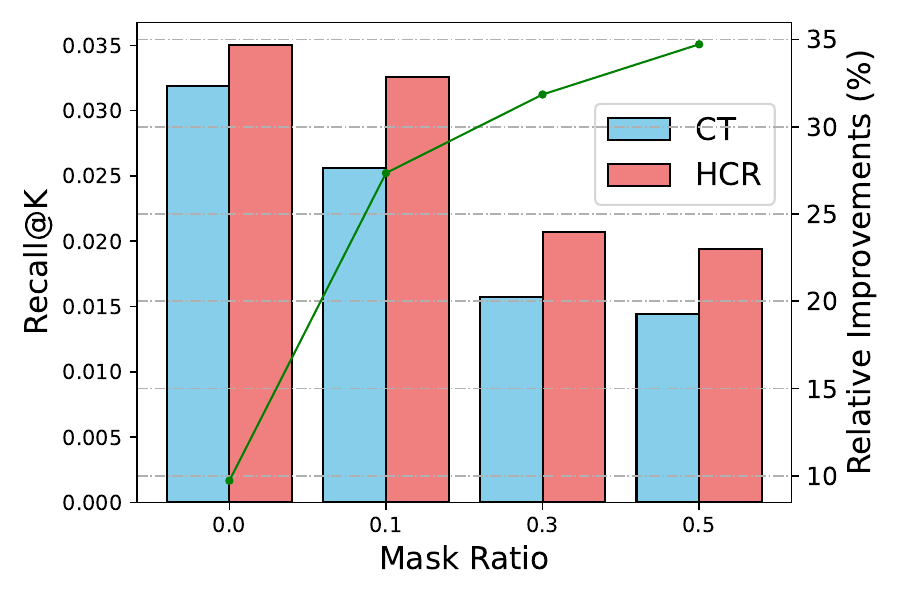}
	}
	\subfigure[Absolute performance of HCR and CT and relative drop of HCR and CT compared with the original performance.]{
		\includegraphics[width=0.45\columnwidth]{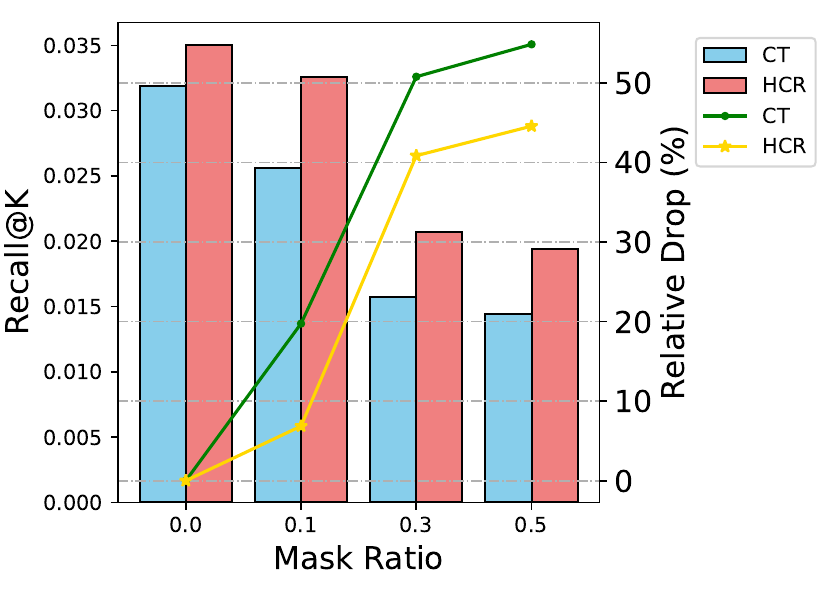}
	}
	\vspace{-0.35cm}
	\caption{Performance of HCR and CT in four datasets with different mask ratios (\ie different strengths of confounding effects) (a) the relative improvements of HCR over CT; (b) the relative drop of HCR and CT.}
	\label{fig:semisimu}
	\vspace{-0.4cm}
\end{figure}

To further investigate source of performance gains of HCR and effects of confounding strength, we generate semi-simulated datasets from Taobao and evaluate HCR and the strongest baseline CT. The semi-simulated datasets are generated by selectively masking a ratio of like feedback according to the item feature. The masking can be viewed as introducing hidden confounders that affect item features and like feedback into the training datasets. The larger ratio means stronger hidden confounding effects. We keep validation and test sets consistent with the original experiments. The results are summarized in Figure \ref{fig:semisimu}.
Figure \ref{fig:semisimu} (a) shows that HCR achieves greater relative improvements over CT as the mask ratio increases. In datasets with stronger confounding effects (\ie higher mask ratios), HCR has a greater advantage over CT. This result can be attributed to that HCR could recognize causal effects of item features on like feedback, while CT suffers more from confounding effects in the training dataset.
According to Figure \ref{fig:semisimu} (b), as the mask ratio in the training dataset increases, the performance of both HCR and CT deteriorates. Possible reasons of performance drop include increased confounding effects and reduction of training data size. 
While HCR cannot entirely eliminate the confounding effects in real-world datasets due to their complexity, it exhibits smaller decline in performance compared to CT. This outcome demonstrates that HCR can retain relatively accurate estimations of user preferences with increasing confounding effects, thereby validating its ability to mitigate hidden confounding effects in practical scenarios.

\section{Related Work}

In this section, we first overview existing work on de-biasing in recommendation, then we specially discuss the work to deal with the bias from the confounder perspective.

\subsection{De-biasing in Recommendation}
Recommender systems encounter various bias issues, including position bias~\cite{PAL,cross-position}, exposure bias~\cite{IpsImplicit,ExpoMF}, and popularity bias~\cite{PDA,bprpc}. Several methods have been developed to address these biases, following different technical approaches. IPW-based methods adjust the data distribution through sample re-weighting~\cite{IPS-ICML2016,Zhang2020LargescaleCA,autodebias,Guo2021EnhancedDR}, but their performance heavily relies on accurate propensity scores and may exhibit high variance~\cite{Saito2020AsymmetricTF}. Doubly Robust (DR) approaches combine IPW with an imputation model~\cite{wang2019doubly,Enhanced-DR,b-DR}, which also depend on precise propensities for optimal performance.However, these methods do not account for hidden confounding effects, and therefore, the accuracy of propensities cannot be guaranteed in our settings. Another alternative to address bias issues is leveraging unbiased data~\cite{autodebias,causalEmb,liu2020general,wang2021combating}. However, obtaining unbiased data often requires causal intervention, which may negatively impact the user experience~\cite{wang2019doubly}. Furthermore, controlling confounders for intervention experiments becomes infeasible when hidden confounders are present. Heuristic methods like re-ranking~\cite{bprpc,pop-rank} and loss constraints~\cite{chen2020esam,pop-loss} have been proposed to mitigate recommendation bias. However, these methods often lack a solid theoretical foundation and struggle to handle hidden confounders.

\subsection{Deconfounding in Recommendation}
Recently, causality-aware methods have thrived in recommendations, mostly focusing on observed confounders, while a few works focus on hidden confounders. The two types of work are introduced respectively in following.
\subsubsection{Deconfounding for Observed Confounders} 
Recent works utilize causal tools to address bias problems and identify confounding effects as underlying causes~\cite{PDA,wenjie-kdd2021,CauSeR-CIKM2021,sato2020unbiased,IPS-dec,wang2022TKDE}. PDA~\cite{PDA} identifies popularity as a confounder affecting both exposures and clicks. \cite{wenjie-kdd2021} identifies the distribution of historical interactions as a confounding factor amplifying bias. \cite{CauSeR-CIKM2021} identifies the momentum of the SGD optimizer as a confounder, contributing to popularity bias in session-based recommendations. \cite{wang2022TKDE} identifies item aspects (e.g., actor) as confounders in scenarios with diverse information. These works employ backdoor adjustment methods to handle confounding issues, requiring control over confounders.
In contrast, our approach does not identify any specific controllable factor as a confounder, making backdoor adjustment methods unsuitable. \cite{sato2020unbiased} identifies user/item features as confounders when estimating causal effects of recommendation, while \cite{IPS-dec} identifies response rate as a confounder for user satisfaction, both employing IPW for deconfounding. Additionally, counterfactual inference-based methods~\cite{wang_click,wei2021MACR} also tackle confounding bias, assuming the availability of measurable confounders.

\subsubsection{Deconfounding for Hidden Confounders}
DCF~\cite{DCF-recsys20} considers hidden confounders related to exposures and learns exposure models to compute substitutes. 
However, DCF's substitute confounder estimation may not effectively address hidden confounders weakly related to exposures.
 \cite{InfoConfounding} addresses the case where hidden confounders affect treatment and outcomes in model inputs. Biased representations are learned and the biased component is discarded during inference using an information bottleneck approach. However, confounders such as news events may not be included in model inputs. Instead, HCR focuses on general confounders affecting item attributes and user feedback, without assuming that confounding effects are fully reflected in embedding representations.

For sequential recommendation, ~\cite{Wang2022UnbiasedSR} proposes an unbiased approach by modeling hidden confounders. This work achieves de-confounding using IPW with estimated propensities. DEMER \cite{Shang2019EnvironmentRW} focuses on reinforcement learning-based sequential recommendation and uses confounding agents to simulate confounders in the environment reconstruction. 
It introduces a confounder embedded policy and a compatible discriminator for deconfounded environment reconstruction. 
However, adapting these methods to general recommendation settings is challenging. Disentangling hidden confounding effects without sufficient inductive bias or supervised information remains an open problem \cite{zheng2021disentangling}. Instead, HCR identify the causal effect directly without explicitly modeling the hidden confounder.

\section{Conclusion and Future Work}
In this paper, we highlighted the importance of considering hidden confounders in recommender systems. We resorted to causal language to abstract the recommendation process as a simplified causal graph with some assumptions. 
Inspired by the front-door adjustment technique rooted in causality theory, we proposed a novel deconfounded training and inference framework named Hidden Confounder Removal (HCR), which mitigates the hidden confounding effect when estimating the causal effect $P(l|u, do(i))$.
In this work, we focused on a simplified causal model and instantiated HCR in micro-videos and e-commerce product recommendation scenarios over a representative multi-modal model MMGCN. 
Empirical results on three real-world datasets validated the advantages of mitigating hidden confounding effects.

Currently, HCR is based on the assumption of no direct edge from the confounder $V$ to the mediator $M$ in the proposed causal graph, which could experience partial violations in complex real-world situations. Therefore, the applicability of HCR may have limitations in certain scenarios.
In the future, we will extend the proposed HCR framework to handle cases where the mediators are non-existent or do not satisfy the front-door criterion. Besides, we will also consider the following directions:
1) considering hidden confounders between like and user features;
2) investigating the relation between the hidden confounder and bias issues in recommender systems, \eg selection bias and popularity bias; 
and 3) testing HCR framework with more backbone models in recommendation scenarios such as food recommendation. 

\ifCLASSOPTIONcompsoc
  \section*{Acknowledgments}
\else
  \section*{Acknowledgment}
\fi


This work is supported by the National Key Research and Development Program of China (2022YFB3104701), the National Natural Science Foundation of China (62272437, U22A2094, 62121002), and the CCCD Key Lab of Ministry of Culture and Tourism.

\appendices


\ifCLASSOPTIONcaptionsoff
  \newpage
\fi



\small
\bibliographystyle{IEEEtran}
\bibliography{reference}

\begin{thebibliography}{10}
\providecommand{\url}[1]{#1}
\csname url@samestyle\endcsname
\providecommand{\newblock}{\relax}
\providecommand{\bibinfo}[2]{#2}
\providecommand{\BIBentrySTDinterwordspacing}{\spaceskip=0pt\relax}
\providecommand{\BIBentryALTinterwordstretchfactor}{4}
\providecommand{\BIBentryALTinterwordspacing}{\spaceskip=\fontdimen2\font plus
\BIBentryALTinterwordstretchfactor\fontdimen3\font minus \fontdimen4\font\relax}
\providecommand{\BIBforeignlanguage}[2]{{%
\expandafter\ifx\csname l@#1\endcsname\relax
\typeout{** WARNING: IEEEtran.bst: No hyphenation pattern has been}%
\typeout{** loaded for the language `#1'. Using the pattern for}%
\typeout{** the default language instead.}%
\else
\language=\csname l@#1\endcsname
\fi
#2}}
\providecommand{\BIBdecl}{\relax}
\BIBdecl

\bibitem{deep-matching}
J.~Xu, X.~He, and H.~Li, ``Deep learning for matching in search and recommendation,'' \emph{Found. Trends Inf. Retr.}, pp. 102--288, 2020.

\bibitem{SRec-survey}
H.~Fang, D.~Zhang, Y.~Shu, and G.~Guo, ``Deep learning for sequential recommendation: Algorithms, influential factors, and evaluations,'' \emph{ACM Trans. Inf. Syst.}, vol.~39, no.~1, pp. 10:1--10:42, 2020.

\bibitem{chen_bias_2020}
J.~Chen, H.~Dong, X.~Wang, F.~Feng, M.~Wang, and X.~He, ``\BIBforeignlanguage{en}{Bias and {Debias} in {Recommender} {System}: {A} {Survey} and {Future} {Directions}},'' \emph{\BIBforeignlanguage{en}{arXiv:2010.03240}}, 2020.

\bibitem{Saito2020AsymmetricTF}
Y.~Saito, ``Asymmetric tri-training for debiasing missing-not-at-random explicit feedback,'' in \emph{Proceedings of the 43rd International ACM SIGIR Conference on Research and Development in Information Retrieval}, 2020, pp. 309--318.

\bibitem{PDA}
Y.~Zhang, F.~Feng, X.~He, T.~Wei, C.~Song, G.~Ling, and Y.~Zhang, ``Causal intervention for leveraging popularity bias in recommendation,'' in \emph{Proceedings of the 44th International ACM SIGIR Conference on Research and Development in Information Retrieval}, 2021, p. 11–20.

\bibitem{pearl-primer}
J.~Pearl, M.~Glymour, and N.~P. Jewell, \emph{Causal inference in statistics: A primer}.\hskip 1em plus 0.5em minus 0.4em\relax John Wiley \& Sons, 2016.

\bibitem{IPS-ICML2016}
T.~Schnabel, A.~Swaminathan, A.~Singh, N.~Chandak, and T.~Joachims, ``Recommendations as treatments: Debiasing learning and evaluation,'' in \emph{International Conference on Machine learning}, 2016, pp. 1670--1679.

\bibitem{Zhang2020LargescaleCA}
W.~Zhang, W.~Bao, X.-Y. Liu, K.~Yang, Q.~Lin, H.~Wen, and R.~Ramezani, ``Large-scale causal approaches to debiasing post-click conversion rate estimation with multi-task learning,'' in \emph{Proceedings of The Web Conference 2020}, 2020, pp. 2775--2781.

\bibitem{Gruson2019OfflineET}
A.~Gruson, P.~Chandar, C.~Charbuillet, J.~McInerney, S.~Hansen, D.~Tardieu, and B.~Carterette, ``Offline evaluation to make decisions about playlistrecommendation algorithms,'' in \emph{Proceedings of the Twelfth ACM International Conference on Web Search and Data Mining}, 2019, pp. 420--428.

\bibitem{autodebias}
J.~Chen, H.~Dong, Y.~Qiu, X.~He, X.~Xin, L.~Chen, G.~Lin, and K.~Yang, ``Autodebias: Learning to debias for recommendation,'' in \emph{Proceedings of the 44th International ACM SIGIR Conference on Research and Development in Information Retrieval}, 2021, p. 21–30.

\bibitem{IpsImplicit}
Y.~Saito, S.~Yaginuma, Y.~Nishino, H.~Sakata, and K.~Nakata, ``Unbiased recommender learning from missing-not-at-random implicit feedback,'' in \emph{Proceedings of the 13th International Conference on Web Search and Data Mining}, 2020, pp. 501--509.

\bibitem{Guo2021EnhancedDR}
S.~Guo, L.~Zou, Y.~Liu, W.~Ye, S.~Cheng, S.~Wang, H.~Chen, D.~Yin, and Y.~Chang, ``Enhanced doubly robust learning for debiasing post-click conversion rate estimation,'' in \emph{Proceedings of the 44th International ACM SIGIR Conference on Research and Development in Information Retrieval}, 2021, pp. 275--284.

\bibitem{wenjie-kdd2021}
W.~Wang, F.~Feng, X.~He, X.~Wang, and T.-S. Chua, ``Deconfounded recommendation for alleviating bias amplification,'' in \emph{Proceedings of the 27th ACM SIGKDD Conference on Knowledge Discovery \& Data Mining}, 2021, p. 1717–1725.

\bibitem{CauSeR-CIKM2021}
P.~Gupta, A.~Sharma, P.~Malhotra, L.~Vig, and G.~Shroff, ``Causer: Causal session-based recommendations for handling popularity bias,'' in \emph{Proceedings of the 30th ACM International Conference on Information \& Knowledge Management}, 2021, pp. 3048--3052.

\bibitem{CausalAttention}
J.~Zhang, X.~Chen, and W.~X. Zhao, ``Causally attentive collaborative filtering,'' in \emph{Proceedings of the 30th ACM International Conference on Information \& Knowledge Management}, 2021, pp. 3622--3626.

\bibitem{10.1145/3407190}
Y.~Deldjoo, M.~Schedl, P.~Cremonesi, and G.~Pasi, ``Recommender systems leveraging multimedia content,'' \emph{ACM Comput. Surv.}, vol.~53, no.~5, sep 2020.

\bibitem{MMGCN}
Y.~Wei, X.~Wang, L.~Nie, X.~He, R.~Hong, and T.-S. Chua, ``Mmgcn: Multi-modal graph convolution network for personalized recommendation of micro-video,'' in \emph{Proceedings of the 27th ACM International Conference on Multimedia}, 2019, pp. 1437--1445.

\bibitem{Wu_survey_neural}
L.~Wu, X.~He, X.~Wang, K.~Zhang, and M.~Wang, ``A survey on neural recommendation: From collaborative filtering to content and context enriched recommendation,'' \emph{arXiv preprint arXiv:2104.13030}, 2021.

\bibitem{gao2019learning}
C.~Gao, X.~He, D.~Gan, X.~Chen, F.~Feng, Y.~Li, T.-S. Chua, L.~Yao, Y.~Song, and D.~Jin, ``Learning to recommend with multiple cascading behaviors,'' \emph{IEEE Transactions on Knowledge and Data Engineering}, 2019.

\bibitem{wang_click}
W.~Wang, F.~Feng, X.~He, H.~Zhang, and T.-S. Chua, ``Clicks can be cheating: Counterfactual recommendation for mitigating clickbait issue,'' in \emph{Proceedings of the 44th International ACM SIGIR Conference on Research and Development in Information Retrieval}, 2021, p. 1288–1297.

\bibitem{MTL_survey}
Y.~Zhang and Q.~Yang, ``A survey on multi-task learning,'' \emph{IEEE Transactions on Knowledge and Data Engineering}, pp. 1--1, 2021.

\bibitem{ma_entire_2018}
X.~Ma, L.~Zhao, G.~Huang, Z.~Wang, Z.~Hu, X.~Zhu, and K.~Gai, ``Entire space multi-task model: An effective approach for estimating post-click conversion rate,'' in \emph{Proceedings of the 41st International ACM SIGIR Conference on Research and Development in Information Retrieval}, 2018, pp. 1137--1140.

\bibitem{DCF-recsys20}
Y.~Wang, D.~Liang, L.~Charlin, and D.~M. Blei, ``Causal inference for recommender systems,'' in \emph{14th ACM Conference on Recommender Systems}, 2020, pp. 426--431.

\bibitem{kingma_adam_2017}
D.~P. Kingma and J.~Ba, ``Adam: {A} method for stochastic optimization,'' in \emph{3rd International Conference on Learning Representations}, 2015.

\bibitem{PAL}
H.~Guo, J.~Yu, Q.~Liu, R.~Tang, and Y.~Zhang, ``Pal: a position-bias aware learning framework for ctr prediction in live recommender systems,'' in \emph{Proceedings of the 13th ACM Conference on Recommender Systems}, 2019, pp. 452--456.

\bibitem{cross-position}
H.~Zhuang, Z.~Qin, X.~Wang, M.~Bendersky, X.~Qian, P.~Hu, and D.~C. Chen, ``Cross-positional attention for debiasing clicks,'' in \emph{Proceedings of the Web Conference 2021}, 2021, pp. 788--797.

\bibitem{ExpoMF}
D.~Liang, L.~Charlin, J.~McInerney, and D.~M. Blei, ``Modeling user exposure in recommendation,'' in \emph{Proceedings of the 25th international conference on World Wide Web}, 2016, pp. 951--961.

\bibitem{bprpc}
Z.~Zhu, Y.~He, X.~Zhao, Y.~Zhang, J.~Wang, and J.~Caverlee, ``Popularity-opportunity bias in collaborative filtering,'' in \emph{Proceedings of the 14th ACM International Conference on Web Search and Data Mining}, 2021, pp. 85--93.

\bibitem{wang2019doubly}
X.~Wang, R.~Zhang, Y.~Sun, and J.~Qi, ``Doubly robust joint learning for recommendation on data missing not at random,'' in \emph{International Conference on Machine Learning}, 2019, pp. 6638--6647.

\bibitem{Enhanced-DR}
S.~Guo, L.~Zou, Y.~Liu, W.~Ye, S.~Cheng, S.~Wang, H.~Chen, D.~Yin, and Y.~Chang, ``Enhanced doubly robust learning for debiasing post-click conversion rate estimation,'' in \emph{Proceedings of the 44th International ACM SIGIR Conference on Research and Development in Information Retrieval}, 2021, p. 275–284.

\bibitem{b-DR}
A.~Gilotte, C.~Calauz{\`e}nes, T.~Nedelec, A.~Abraham, and S.~Doll{\'e}, ``Offline a/b testing for recommender systems,'' in \emph{Proceedings of the 11th ACM International Conference on Web Search and Data Mining}, 2018, pp. 198--206.

\bibitem{causalEmb}
S.~Bonner and F.~Vasile, ``Causal embeddings for recommendation,'' in \emph{Proceedings of the 12th ACM conference on recommender systems}, 2018, pp. 104--112.

\bibitem{liu2020general}
D.~Liu, P.~Cheng, Z.~Dong, X.~He, W.~Pan, and Z.~Ming, ``A general knowledge distillation framework for counterfactual recommendation via uniform data,'' in \emph{Proceedings of the 43rd International ACM SIGIR Conference on Research and Development in Information Retrieval}, 2020, pp. 831--840.

\bibitem{wang2021combating}
X.~Wang, R.~Zhang, Y.~Sun, and J.~Qi, ``Combating selection biases in recommender systems with a few unbiased ratings,'' in \emph{Proceedings of the 14th ACM International Conference on Web Search and Data Mining}, 2021, pp. 427--435.

\bibitem{pop-rank}
H.~Abdollahpouri, R.~Burke, and B.~Mobasher, ``Managing popularity bias in recommender systems with personalized re-ranking,'' in \emph{Proceedings of the Thirty-Second International Florida Artificial Intelligence Research Society Conference}, 2019, pp. 413--418.

\bibitem{chen2020esam}
Z.~Chen, R.~Xiao, C.~Li, G.~Ye, H.~Sun, and H.~Deng, ``Esam: Discriminative domain adaptation with non-displayed items to improve long-tail performance,'' in \emph{Proceedings of the 43rd International ACM SIGIR Conference on Research and Development in Information Retrieval}, 2020, pp. 579--588.

\bibitem{pop-loss}
H.~Abdollahpouri, R.~Burke, and B.~Mobasher, ``Controlling popularity bias in learning-to-rank recommendation,'' in \emph{Proceedings of the Eleventh ACM Conference on Eecommender Systems}, 2017, pp. 42--46.

\bibitem{sato2020unbiased}
M.~Sato, S.~Takemori, J.~Singh, and T.~Ohkuma, ``Unbiased learning for the causal effect of recommendation,'' in \emph{Fourteenth ACM Conference on Recommender Systems}, 2020, pp. 378--387.

\bibitem{IPS-dec}
K.~Christakopoulou, M.~Traverse, T.~Potter, E.~Marriott, D.~Li, C.~Haulk, E.~H. Chi, and M.~Chen, ``Deconfounding user satisfaction estimation from response rate bias,'' in \emph{Fourteenth ACM Conference on Recommender Systems}, 2020, pp. 450--455.

\bibitem{wang2022TKDE}
X.~Wang, Q.~Li, D.~Yu, P.~Cui, Z.~Wang, and G.~Xu, ``Causal disentanglement for semantics-aware intent learning in recommendation,'' \emph{IEEE Transactions on Knowledge and Data Engineering}, 2022.

\bibitem{wei2021MACR}
T.~Wei, F.~Feng, J.~Chen, Z.~Wu, J.~Yi, and X.~He, ``Model-agnostic counterfactual reasoning for eliminating popularity bias in recommender system,'' in \emph{Proceedings of the 27th ACM SIGKDD Conference on Knowledge Discovery \& Data Mining}, 2021, pp. 1791--1800.

\bibitem{InfoConfounding}
D.~Liu, P.~Cheng, H.~Zhu, Z.~Dong, X.~He, W.~Pan, and Z.~Ming, ``Mitigating confounding bias in recommendation via information bottleneck,'' in \emph{Fifteenth ACM Conference on Recommender Systems}, 2021, p. 351–360.

\bibitem{Wang2022UnbiasedSR}
Z.~Wang, S.~Shen, Z.~Wang, B.~Chen, X.~Chen, and J.-R. Wen, ``Unbiased sequential recommendation with latent confounders,'' in \emph{Proceedings of the ACM Web Conference 2022}, 2022, pp. 2195--2204.

\bibitem{Shang2019EnvironmentRW}
W.~Shang, Y.~Yu, Q.~Li, Z.~Qin, Y.~Meng, and J.~Ye, ``Environment reconstruction with hidden confounders for reinforcement learning based recommendation,'' in \emph{Proceedings of the 25th ACM SIGKDD International Conference on Knowledge Discovery \& Data Mining}, 2019, pp. 566--576.

\bibitem{zheng2021disentangling}
Y.~Zheng, C.~Gao, X.~Li, X.~He, Y.~Li, and D.~Jin, ``Disentangling user interest and conformity for recommendation with causal embedding,'' in \emph{Proceedings of the Web Conference 2021}, 2021, pp. 2980--2991.

\end{thebibliography}

\vspace{-1cm}

\end{document}